\documentclass[11pt]{article}
\usepackage{amsmath,amssymb,amsthm} 
\usepackage{bbm}
\usepackage{stmaryrd}
\usepackage{graphicx}
\usepackage{paralist}
\setdefaultenum{(i)}{}{}{}
\usepackage[margin=1in]{geometry}
\usepackage[round]{natbib}
\defcitealias{gerhold.al.11}{GGMKS}

\newcommand\ve{\varepsilon}

\DeclareMathAlphabet\scr{U}{scr}{m}{n}
\SetMathAlphabet\scr{bold}{U}{scr}{b}{n}
  \DeclareFontFamily{U}{scr}{\skewchar\font'177}%
  \DeclareFontShape{U}{scr}{m}{n}{<-6>rsfs5<6-8>rsfs7<8->rsfs10}{}%
  \DeclareFontShape{U}{scr}{b}{n}{<-6>rsfs5<6-8>rsfs7<8->rsfs10}{}%

\newtheorem{theorem}{Theorem}[section]

\newtheorem{definition}[theorem]{Definition}
\newtheorem{lemma}[theorem]{Lemma}
\newtheorem{proposition}[theorem]{Proposition}

\theoremstyle{definition}

\newcommand{\esp}[2][E]{#1\left[#2\right]}

\newcommand{\eps}{\varepsilon}
\newcommand{\nada}[1]{}

\numberwithin{equation}{section}

\renewenvironment{thebibliography}[1]{%
\begin{oldthebibliography}{#1}%
\setlength{\baselineskip}{.9em}
\linespread{.9}
\small
\setlength{\parskip}{0ex}%
\setlength{\itemsep}{.1em}%
}%
{%
\end{oldthebibliography}%
}

\DeclareMathOperator{\lip}{LiPr}
\DeclareMathOperator{\sht}{ShTu}
\DeclareMathOperator{\wet}{WeTu}
\DeclareMathOperator{\cer}{EA}
\DeclareMathOperator{\esr}{ESR}
\DeclareMathOperator{\prof}{Profit}

\begin{document}

\title{Long Horizons, High Risk-Aversion, and Endogenous Spreads\footnote{We thank Ales \v{C}ern{\'y}, Stefan Gerhold, Marcel Nutz, and Walter Schachermayer for helpful discussions. We also thank two anonymous referees and Ren Liu for their careful reading of the paper. Part of this work was completed while the second author was visiting Columbia University. He thanks Ioannis Karatzas and the university for their hospitality.}}
\author{Paolo Guasoni\thanks{Boston University, Department of Mathematics and Statistics, 111 Cummington Street Boston, MA 02215, USA.
Dublin City University, School of Mathematical Sciences, Glasnevin, Dublin 9, Ireland, email: \texttt{guasoni@bu.edu}. Partially supported by the ERC (278295), NSF (DMS-0807994, DMS-1109047), SFI (07/MI/008, 07/SK/M1189, 08/SRC/FMC1389), and FP7 (RG-248896).}
\and
Johannes Muhle-Karbe\thanks{ETH Z\"urich, Departement Mathematik, R\"amistrasse 101, CH-8092, Z\"urich, Switzerland, and Swiss Finance Institute, email:
\texttt{johannes.muhle-karbe@math.ethz.ch}. Partially supported by the National Centre of Competence in Research ``Financial Valuation and Risk Management'' (NCCR FINRISK), Project D1 (Mathematical Methods in Financial Risk Management), of the Swiss National Science Foundation (SNF).}}

\maketitle

\begin{abstract}
For an investor with constant absolute risk aversion and a long horizon, who trades in a market with constant investment opportunities and small proportional transaction costs, we obtain explicitly the optimal investment policy, its implied welfare, liquidity premium, and trading volume. We identify these quantities as the limits of their isoelastic counterparts for high levels of risk aversion. The results are robust with respect to finite horizons, and extend to multiple uncorrelated risky assets.

In this setting, we study a Stackelberg equilibrium, led by a risk-neutral, monopolistic market maker who sets the spread as to maximize profits. The resulting endogenous spread depends on investment opportunities only, and is of the order of a few percentage points for realistic parameter values.
\end{abstract}




\newpage
\section{Introduction}

Despite their singular behavior, investors with constant {absolute} risk aversion are familiar figures in financial economics, thanks to their tractable character. Such investors, defined by exponential utility functions, are indeed peculiar for both portfolios and prices. With constant investment opportunities, they insist on keeping in risky assets a fixed amount of money, regardless of their wealth. If asked to price a claim, their answer depends neither on wealth, nor on risk aversion. Consuming over time, they do not disdain negative consumption, especially at later dates.
Yet, their actions are often easier to grasp than the proper but impenetrable behavior of isoelastic investors.\footnote{An isoelastic investor is one with constant \emph{relative} risk aversion, i.e. with either power of logarithmic utility.} Hence, exponential utility remains a central tool to glean insights from complex models, such as the ones with frictions.

This paper examines the implications of exponential utility for long-run portfolio choice with small transaction costs. For an exponential investor, with constant investment opportunities and a long planning horizon, we find the optimal trading policy, its welfare, the liquidity premium, and trading volume. We then allow a risk neutral, monopolistic market maker to set the spread as to maximize profits, obtaining an endogenous spread that depends on investment opportunities alone.

Our analysis leads to new economic implications, and sheds new light on existing results.
We find that investing optimally on a long horizon is equivalent to receiving, over the same period, a fixed \emph{equivalent annuity}, found explicitly, which does not depend on the horizon. This fact is well-known in the frictionless case, but fails for transaction costs with a finite horizon, due to the spurious effects of portfolio set-up and liquidation. As in the isoelastic case, transaction costs entail a small reduction in the equivalent annuity.

The equivalent annuity, trading boundaries, and absolute turnover are inversely proportional to risk aversion, as in the frictionless case, while relative turnover and the liquidity premium are independent of risk aversion.
We identify all these quantities as the limits of their isoelastic counterparts, as relative risk aversion becomes large. This result suggests that exponential utility is a useful tool to study isoelastic investors with high risk aversion. 

Our results are robust to finite horizons and to several assets.
For a finite horizon, we derive bounds on the investor's certainty equivalent, whose time average converges to the equivalent annuity. In a market with several uncorrelated assets, one-dimensional trading policies remain optimal, leading to the same liquidity premia and trading volumes, while equivalent annuities add in the cross section.

Finally, we endogenize the spread by allowing a risk-neutral, monopolistic market maker to fix it to maximize expected profits. Unlike the investor, whose policy and welfare depend on the bid and ask prices alone, the market maker's profits depend on the book price, which lies within the bid-ask spread, and represents the price at which the market maker's inventory is valued. The resulting endogenous spread is independent of risk aversion, and hence depends only on investment opportunities, and on the book price. When the latter is chosen close to the ask price, realistic values of investment opportunities lead to spreads of a few percentage points.

Our results also have novel mathematical implications. We obtain new finite-horizon bounds, which measure the monetary value of investment opportunities for an exponential investor, and are expressed in terms of the risk neutral probability. For isoelastic investors, such bounds involve the myopic probability, under which a hypothetical logarithmic investor adopts the same policy as the original investor under the physical probability. Thus, for exponential investors, the risk neutral probability plays a similar role as the myopic probability for isoelastic investors. This analogy is central to obtain a new kind of verification theorem, which stems from the finite-horizon bounds.

Our paper touches on three main strands of literature: asymptotics, shadow prices, and exponential utility. Our contribution to asmptotics is to prove the first rigorous expansions for small transaction costs and exponential utility, complementing the heuristics of  \citet*{whalley.wilmott.97}, \citet*{atkinson.mokkhaseva.02}, and \citet*{goodman.ostrov.10}, as well as results for the isoelastic case, cf.\ \citet*{MR1284980}, \citet*{rogers.01}, \citet*{MR2048827}, \citet*{gerhold.al.10a,gerhold.al.10b}, and \citet*{bichuch.11}. Our solution is based on shadow prices, as in the recent papers \citet*{MR2676941,gerhold.al.10a,gerhold.al.10b, gerhold.al.11, herczegh2011shadow, choi2012shadow}.

More broadly, our results are relevant for the literature on exponential utility with transaction costs, both in the context of portfolio choice \citep*{atkinson.mokkhaseva.02,liu.04,goodman.ostrov.10}
and of option pricing \citep*{davis.al.93,whalley.wilmott.97,barles.soner.98}.
In contrast to these papers, we remove consumption and random endowment from our model, focusing instead on long-horizon asymptotics for tractability. This approach is common for isoelastic utilities \citep*{MR942619,dumas.luciano.91,gerhold.al.10a}. Unaccountably, it seems unexplored in the exponential class. Our analysis of endogenous spreads is closest to the work of \citet*{luciano2011} on equilibrium between isoelastic investors and dealers, with the difference that we pair an exponential investor with a risk-neutral dealer.

Finally, this paper on constant \emph{absolute} risk aversion complements the analysis of constant \emph{relative} risk aversion in \citet*{gerhold.al.11} (henceforth \citetalias{gerhold.al.11}). To facilitate comparison, the main results in both papers are stated in the same format and with the same notation. Yet, each paper addresses a different set of implications, which is specific to the preference class considered.

The rest of the article is organized as follows. Section 2 presents the model and our main results. Their main implications are discussed in Section 3. The main results are first derived informally in Section 4, then proved in Section 5.

\section{Model and Main Result}

\subsection{Market}
The market has a safe asset $S^0_t=1$ and a risky asset, trading at ask (buying) price $S$. The bid (selling) price is $S(1-\ve)$, hence $\ve \in (0,1)$ is the relative bid-ask spread. Denoting by $W$ a standard Brownian motion, the ask price $S$ follows \begin{equation*}
dS_t/S_t=\mu dt+\sigma dW_t,
\end{equation*}
where $\mu>0$ is the expected excess return and $\sigma>0$ is the volatility. The \emph{mean-variance ratio} $\bar\mu=\mu/\sigma^2$ turns out to be a key parameter in the solution.

A self-financing \emph{trading strategy} is an $\mathbb{R}^2$-valued predictable process $(\varphi^0,\varphi)$ of finite variation: $(\varphi^0_{0-},\varphi_{0-}) = (\xi^0,\xi) \in \mathbb{R}^2$ represents the initial positions (in units) in the safe and risky asset, and $\varphi^0_t$ and $\varphi_t$ denote the positions held at time $t \geq 0$. 
Writing $\varphi_t = \varphi^{\uparrow}_t-\varphi^{\downarrow}_t$ as the difference between the cumulative number of shares bought ($\varphi^{\uparrow}_t$) and sold ($\varphi^{\downarrow}_t$) by time $t$, the \emph{self-financing condition} dictates that the cash balance $\varphi^0$ changes only due to trading activity in the number of shares $\varphi$:
\begin{equation*}\label{eq:selffinancing}
d\varphi^0_t =  -S_t d\varphi_t^{\uparrow}+ (1-\ve)S_t d\varphi^{\downarrow}_t .
\end{equation*}
Trading strategies are further restricted from unlimited borrowing by the following admissibility condition, which rules out doubling strategies:\footnote{The definition of admissibility given here is sufficient to guarantee an interior solution in our setting, and has a clear interpretation. In more abstract models, admissibility is typically defined in terms of pricing measures, cf.\  \citet{delbaen.al.02,kabanov.stricker.02,schachermayer.03}.}
\begin{definition}\label{defi:admissible}
A self-financing strategy $(\varphi^0,\varphi)$ is \emph{admissible} if the risky position $\varphi S$ is a.s. uniformly bounded.
\end{definition}
The \emph{liquidation value} of the wealth associated to an admissible strategy is denoted by
$$\Xi_T^\varphi=\varphi^0_T+\varphi_T^+ (1-\ve)S_T-\varphi_T^- S_T .$$

\subsection{Preferences}

An investor with constant absolute risk aversion $\alpha>0$, which corresponds to the exponential utility function $U(x)=-e^{-\alpha x}$, maximizes the certainty equivalent $U^{-1}(E[U(\Xi)])$, which for exponential utility reduces to:
\begin{equation*}
-\frac{1}{\alpha}\log \esp{e^{-\alpha \Xi_T^\varphi}}.
\end{equation*}
In a market with constant investment opportunities, the certainty equivalent increases linearly with the horizon. Hence, we focus on the certainty equivalent per unit of time, which is the same as a fixed annuity.
\begin{definition}
A strategy $(\varphi^0,\varphi)$ is \emph{long-run optimal} if it maximizes the \emph{equivalent annuity}
 \begin{equation}\label{eq:longrun}
\cer_\alpha^\varphi=\liminf_{T \to \infty} -\frac{1}{\alpha T}\log
E\left[e^{-\alpha \Xi_T^\varphi}\right].
 \end{equation}
$\cer_\alpha=\max_\varphi \cer_\alpha^\varphi$ denotes the \emph{maximal equivalent annuity}.
\end{definition}

\subsection{Main Result}

The next theorem contains our main results. Recall that $\bar\mu=\mu/\sigma^2$ is the mean-variance ratio.

\begin{theorem}
\label{thm:mainresult}
An investor with constant absolute risk aversion $\alpha>0$ trades to maximize the equivalent annuity. Then, for a small bid-ask spread $\varepsilon>0$:
\begin{enumerate}[i)]
\item 
\emph{(Equivalent Annuity)}\\
For the investor, trading the risky asset with transaction costs is equivalent to leaving all wealth in the safe asset, while receiving the \emph{equivalent annuity} (the \emph{gap} $\bar\lambda$ is defined in $iv)$ below):
\begin{equation*}
\cer_\alpha=\sigma^2\bar\beta=\frac{\sigma^2}{2\alpha} (\bar\mu^2-\bar\lambda^2).
\end{equation*}

\item
\emph{(Liquidity Premium)}\\
Trading the risky asset with transaction costs is equivalent to trading a hypothetical asset, at no transaction costs, with the same volatility $\sigma$, but with a lower expected excess return $\sigma^2\sqrt{\bar{\mu}^2-\bar\lambda^2}$. Thus, the \emph{liquidity premium} is
\begin{equation*}
\lip = \sigma^2\left(\bar\mu-\sqrt{\bar\mu^2-\bar\lambda^2}\right).
\end{equation*}

\item 
\emph{(Trading Policy)}\\
It is optimal to keep the value of the risky position within the buy and sell boundaries
\begin{equation*}\label{portfolios}
\eta_{\alpha-}=\frac{\bar\mu-\bar\lambda}{\alpha},
\qquad
\eta_{\alpha+}=\frac{\bar\mu+\bar\lambda}{\alpha},
\end{equation*}
where $\eta_{\alpha-}$ and $\eta_{\alpha+}$ are evaluated at ask and bid prices, respectively.
\item 
\emph{(Gap)}\\
The \emph{gap} $\bar\lambda$ is the unique value for which the solution of the initial value problem
\begin{align*}
& w'(y)-w(y)^2+\left(2\bar\mu-1\right)w(y)-(\bar\mu-\bar\lambda)(\bar\mu+\bar\lambda)= 0\\
& w(0) = \bar\mu-\bar\lambda,
\end{align*}
also satisfies the terminal value condition
\begin{equation*}
w\left(\log \left(\frac{u(\bar\lambda)}{l(\bar\lambda)}\right)\right) = \bar\mu+\bar\lambda,
\qquad\text{where}\qquad
u(\bar\lambda)=\frac{\bar\mu+\bar\lambda}{(1-\ve)\alpha}\quad \mbox{and}\quad l(\bar\lambda)=\frac{\bar\mu-\bar\lambda}{\alpha}.
\end{equation*}
In view of the explicit formula for $w(y,\bar\lambda)$ in Lemma~\ref{lem:riccati} below, this is a scalar equation for $\bar\lambda$.
\item
\emph{(Trading Volume)}\\
Let $\mu \neq \sigma^2/2$.\footnote{The special case $\mu=\sigma^2/2$ leads to analogous results, see GGMKS, Lemma D.2.} Then, \emph{relative turnover}, defined as shares traded $d||\varphi||_t$ divided by shares held $|\varphi_t|$, has long-term average
\begin{align*}
\sht =
\lim_{T\rightarrow\infty}\frac1T\int_0^T \frac{d\|\varphi\|_t}{|\varphi_t|}=
\frac{\sigma^2}{2}\left(\frac{1-2\bar\mu}{(u(\bar\lambda)/l(\bar\lambda))^{1-2\bar\mu}-1}+\frac{2\bar\mu-1}{(u(\bar\lambda)/l(\bar\lambda))^{2\bar\mu-1}-1}\right).
\end{align*}
\emph{Absolute turnover}, defined as value of wealth traded, has long term-average:\footnote{The number of shares is written as the difference $\varphi_t=\varphi^{\uparrow}_t-\varphi^{\downarrow}_t$ of cumulative shares bought and sold, and wealth is evaluated at trading prices, i.e., at the bid price $(1-\ve)S_t$ when selling, and at the ask price $S_t$ when buying.} 
\begin{align*}
\wet_\alpha &= \lim_{T\to \infty} \frac{1}{T}\left(\int_0^T (1-\ve)S_t d\varphi^{\downarrow}_t+\int_0^T S_t d\varphi^{\uparrow}_t\right)\\
&= \frac{\sigma^2}{2}\left(\frac{\eta_{\alpha+}(1-2\bar\mu)}{(u(\bar\lambda)/l(\bar\lambda))^{1-2\bar\mu}-1}+\frac{\eta_{\alpha-}(2\bar\mu-1)}{(u(\bar\lambda)/l(\bar\lambda))^{2\bar\mu-1}-1}\right).
\end{align*}
\item
\emph{(Asymptotics)}\\
The following expansions in terms of the bid-ask spread $\eps$ hold:\footnote{Algorithmic calculations can deliver terms of arbitrarily high order.}
\begin{align*}
\bar\lambda &= 
\left(\frac{3}{4} \bar\mu^2\right)^{1/3} \ve^{1/3}
+ O(\ve).
\\
\cer_\alpha &=
\frac{\sigma^2}{2\alpha}\left(\bar\mu^2-\left(\frac{3}{4}\bar\mu^2\right)^{2/3}\ve^{2/3}+O(\ve^{4/3})\right).\\
\lip &= 
\frac{\sigma^2}{2\bar\mu}\left(\frac{3}{3}\bar\mu^2\right)^{2/3} \ve^{2/3}+O(\ve^{4/3}).\\
\eta_{\alpha\pm} &=\frac{1}{\alpha}\left(\bar{\mu}\pm \left(\frac{3}{4}\bar\mu^2\right)^{1/3}\ve^{1/3}+O(\ve)\right).\\
\sht &= 
\frac{\sigma^2}{2} \bar\mu \left(\frac{3}{4}\bar\mu^2\right)^{-1/3} \ve^{-1/3}+O(\ve^{1/3}).
\end{align*}
\newpage
\begin{align*}
\wet_\alpha &= 
\frac{2\sigma^2}{3\alpha}\left(\frac{3}{4}\bar\mu^2\right)^{2/3}\ve^{-1/3}+O(\ve^{1/3}).
\end{align*}
\end{enumerate}
\end{theorem}

The proof of Theorem \ref{thm:mainresult} exploits the construction of a shadow price, i.e., a fictitious risky asset evolving within the bid-ask spread, which is equivalent to the transaction cost market in terms of both welfare \emph{and} the optimal policy. This is the approach used for power utility by  \citet*{gerhold.al.11}.

\begin{theorem}[Shadow price]\label{thm:shadow}
The policy in Theorem \ref{thm:mainresult} $iii)$ and the equivalent annuity in Theorem \ref{thm:mainresult} $i)$ are also optimal for a frictionless risky asset with \emph{shadow price} $\tilde{S}$, which always lies within the bid-ask spread, and coincides with the trading price at times of trading for the optimal policy. The shadow price follows
\begin{equation*}
d\tilde S_t/\tilde S_t = \tilde{\mu}(\Upsilon_t)dt + \tilde{\sigma}(\Upsilon_t)dW_t.
\end{equation*}
for deterministic functions $\tilde{\mu}(\cdot)$ and $\tilde{\sigma}(\cdot)$ given explicitly in Lemma \ref{lem:dynamics}. The state variable $\Upsilon_t=\log(\varphi_t S_t/l(\bar\lambda))$ represents the centered logarithm of the risky position, which is a Brownian motion with drift reflected to remain in the interval $[0,\log(u(\bar\lambda)/l(\bar\lambda))]$, i.e.,
\begin{equation*}
d\Upsilon_t=(\mu-\sigma^2/2) dt+\sigma dW_t +dL_t-dU_t.
\end{equation*}
\end{theorem}
Here, $L_t$ and $U_t$ are nondecreasing processes, corresponding to the relative purchases and sales, respectively (cf.\ Lemma \ref{lem:strategy}).\footnote{The increasing processes $L_t$ and $U_t$ are explicitly identified by the double Skorokhod map in a finite interval, see \cite*{MR2349573}.}
In the interior of the no-trade region, i.e., when the risky position lies in $(l(\bar\lambda),u(\bar\lambda))$, the numbers of units of the safe and risky asset are constant, and the state variable $\Upsilon_t$ follows Brownian motion with drift. When $\Upsilon_t$ reaches the boundary of the no-trade region, buying or selling takes place so as to keep it within $[l(\lambda),u(\lambda)]$.

\section{Implications and Applications}

Theorem \ref{thm:mainresult} presents both analogies with and departures from the isoelastic case in \citetalias{gerhold.al.11}.

One analogy is that the trading policy depends on the market only through the mean-variance ratio $\bar\mu$. All other quantities also only depend on $\bar\mu$ in \emph{business} time, i.e., when measured with respect to the clock $\tau = \sigma^2 t$ running at the speed of the market's variance. Measured in usual calendar time $t$, they scale linearly with the market's variance $\sigma^2$.

In a departure from power utility, the gap $\bar{\lambda}$ is independent of the investor's risk aversion $\alpha$ here. Hence, the liquidity premium and relative turnover are also common to all investors with constant absolute risk aversion. The equivalent annuity and trading boundaries, however, are inversely proportional to risk aversion, as in the frictionless case, and so is absolute turnover.

Further, in this model a strategy of full investment in the risky asset is never optimal, regardless of the risk aversion $\alpha$, and the only solution without trading obtains with a null risk premium $\bar\mu=0$. The absence of full risky investment is a consequence of constant trading boundaries in terms of monetary amounts rather than fractions of wealth.

In spite of these differences, the model- and preference-free relationships for power utilities \citepalias[Section 3.5]{gerhold.al.11} carry over to the exponential case. For example, the universal relation
$$\lip \approx \frac{3}{4} \varepsilon \sht,$$
between the liquidity premium $\lip$, the spread $\varepsilon$, and the relative turnover $\sht$ remains valid for exponential utilities. Thus, this relation not only holds both across market and preference parameters, but is also the same for both families.

\subsection{Trading Policy and Average Risky Position.}

In analogy with the isoelastic case, the optimal policy is to keep the risky position between two boundaries. Unlike the isoelastic case, but as in the frictionless setting, trading boundaries are measured not as fractions of wealth, but as monetary amounts. The novelty is that these boundaries are symmetric around the frictionless optimum. In apparent contrast,  \citet*{liu.04} finds numerically that average risky holdings (over the investment period) increase with transaction costs, suggesting that more transaction costs reduce an investor's \emph{effective} risk aversion. 
In fact, our results are consistent with his findings, once we observe that, if the risk premium is large enough, the risky position is on average closer to the sell than to the buy boundary, even as the boundaries are equidistant from the frictionless solution at the order $\varepsilon^{2/3}$. To see this, it suffices to calculate the average risky position, using the stationary distribution of the reflected geometric Brownian motion $Y_{t}=\varphi_t S_t=l e^{\Upsilon_t}$:
\begin{equation*}
\lim_{T\rightarrow\infty} \frac1T\int_0^T Y_{t} dt = 
\int_{0}^{\log u/l} l e^y 
\frac{2\bar\mu-1}{(u(\bar\lambda)/l(\bar\lambda))^{2\bar\mu-1}-1}e^{(2\bar\mu-1)y} dy =
\frac{\bar\mu-\frac12}{\bar\mu}
\frac{(u(\bar\lambda)/l(\bar\lambda))^{2\bar\mu}-1}{(u(\bar\lambda)/l(\bar\lambda))^{2\bar\mu-1}-1}.
\end{equation*}
To obtain an  asymptotic expansion for small transaction costs, recall that $l(\bar\lambda)=(\bar\mu-\bar\lambda)/\alpha$ and $u(\bar\lambda)=(\bar\mu+\bar\lambda)/(1-\ve)\alpha$. Then, the expansion for $\bar\lambda$ yields:
\begin{equation}\label{eq:avhold}
\lim_{T\rightarrow\infty} \frac1T\int_0^T Y_t dt = \frac{\bar\mu}{\alpha}\left(1+\frac{\bar\mu-1}{6^{1/3}\bar\mu^{2/3}}\eps^{2/3}+O(\eps)\right).
\end{equation}
As a result, for $\bar\mu>1$ the average risky position tends to be higher than the frictionless value $\bar\mu/\alpha$, and vice versa for $\bar\mu<1$. This effect is entirely due to the skewness of the stationary distribution towards the upper boundary, because the boundaries $(\bar\mu\pm\lambda)/\alpha$ are symmetric around $\bar\mu/\alpha$ at the order $\varepsilon^{2/3}$.
Thus, the estimator $\hat\alpha$ obtained by comparing the average risky holding in \eqref{eq:avhold} to the frictionless formula $\bar\mu/\hat\alpha$ is given by:
\begin{equation}
\hat\alpha = 
\alpha \left(1-\frac{\bar\mu-1}{6^{1/3}\bar\mu^{2/3}}\eps^{2/3}+O(\eps)\right)
\end{equation}
This estimator underestimates true risk aversion for $\bar\mu>1$, with larger transaction costs leading to a larger bias, and explains the observation of \citet*{liu.04} that transaction costs seem to reduce effective risk aversion.

\subsection{Convergence and High Risk Aversion Asymptotics}
The formulas in Theorem \ref{thm:mainresult} for exponential utilities are closely related to the limits of the corresponding isoelastic quantities in \citetalias{gerhold.al.11}. The next result makes this relation precise, and justifies the dual interpretation of the aforementioned formulas as high risk aversion asymptotics for isoelastic investors, in the same spirit as in \citet*{cerny.09} and \citet*{nutz.11}.

\begin{theorem}[High Risk Aversion Asymptotics]\label{thm:raa}
An investor with constant relative risk aversion $\gamma>0$ trades to achieve the maximal equivalent safe rate $\esr_\gamma= \max_\varphi \esr_\gamma(\varphi)$, where 
\begin{equation*}
\esr_\gamma(\varphi) = \lim_{T\rightarrow\infty}\frac{1}{T}\log\esp{(\Xi_T^\varphi)^{1-\gamma}}^{\frac 1{1-\gamma}}.
\end{equation*}
Denote by $\bar\lambda_\gamma$,\footnote{This is the gap in business time, replacing $\mu$ and $\sigma^2$ with $\bar\mu$ and $1$, respectively, in \citetalias[Theorem 2.2]{gerhold.al.11}.} $\lip_\gamma$, $\pi_{\gamma\pm}$, $\sht_\gamma$ and $\wet_\gamma$ the corresponding gap, liquidity premium, trading boundaries (as wealth fractions), share turnover, and (relative) wealth turnover (see \citetalias{gerhold.al.11} for details). Then, as $\gamma\uparrow\infty$, the following properties hold for a small spread $\varepsilon>0$:
\begin{enumerate}
\item The equivalent safe rate times relative risk aversion converges to the equivalent annuity times absolute risk aversion, i.e.,
\begin{equation*}
\lim_{\gamma\uparrow\infty} \gamma\esr_\gamma =
\alpha\cer_\alpha = \frac{\sigma^2}{2} (\bar\mu^2-\bar\lambda^2).
\end{equation*}
 
\item
The gap $\bar\lambda_\gamma$ converges to the gap $\bar\lambda$.

\item 
The liquidity premium $\lip_\gamma$ converges to $\lip$.

\item 
The trading boundaries $\pi_{\gamma\pm}$ as wealth fractions, times $\gamma$, converge to the trading boundaries $\eta_{\alpha\pm}$, as position values, times $\alpha$, i.e.,
\begin{equation*}
\lim_{\gamma\uparrow\infty} \gamma\pi_{\gamma\pm} =
\alpha \eta_{\alpha\pm}. 
\end{equation*}

\item 
Share turnover $\sht_\gamma$ converges to relative turnover $\sht$. 

\item
Relative wealth turnover, times $\gamma$, converges to absolute turnover times $\alpha$.
\end{enumerate}
\end{theorem}
 
This result clarifies some properties of the isoelastic quantities. For example \citepalias[Section 3.4]{gerhold.al.11},  share turnover converges to a finite limit as relative risk aversion increases. Theorem \ref{thm:mainresult} identifies this limit as the relative turnover of exponential utility. By contrast, relative wealth turnover declines to zero as risk aversion increases but, once rescaled by $\gamma$, it converges to absolute turnover for exponential utility.

\subsection{Trading Volume and Endogenous Spreads}

\begin{figure}
\centering
\includegraphics{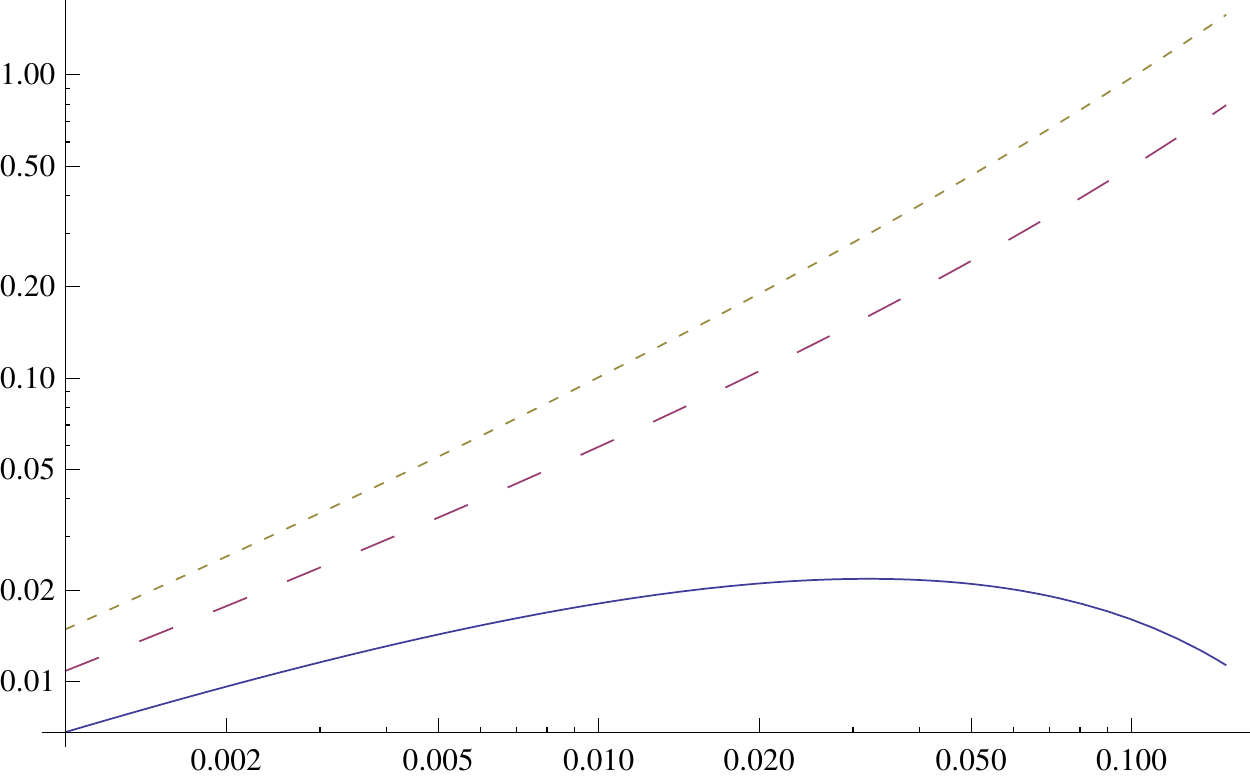}
\caption{\label{fig:plotrans}
Expected value of future fees (vertical axis, in dollars) against the spread $\varepsilon$ (horizontal axis). The plot compares the case of fees earned 100\% on purchases (solid line), 50\% on purchases and sales (long dashing), and 100\% on sales (short dashing).
Parameters are $\mu=8\%, \sigma=16\%$, and $\alpha=(\mu/\sigma^2)/100$, corresponding to a frictionless position of $\eta=\mu/\alpha \sigma^2=100$ dollars, and both axes are in logarithmic scale.}
\end{figure}

An attraction of exponential utility is that, because the value of the investor's risky position is bounded, then also total rebalancing costs are bounded. These costs are in turn related to the profits of a market maker, who earns the costs paid by the investor. Thus, exponential utility is well-suited to develop a valuation model of a market maker, in which both trading and spreads are endogenous. 

If the market maker acts as a monopolist, and fixes the spread to maximize profits, the model yields an endogenous optimal spread, which depends on investment opportunities only. The monopolist tradeoff is clear: a larger spread increases the profit of each transaction, but reduces demand for trading. 

Bid and ask prices alone are not sufficient to determine profits. What is missing is the ``book'' price $\bar S_t$, at which the market maker values his inventory.\footnote{The book price also admits the interpretation of production cost of the asset for the monopolist market maker.} Such a price must lie within the bid-ask spread: whichever policy the investor chooses, the average execution price will be within the bid-ask spread. Thus, we denote the book value by $\bar S_t=S_t(1-\eps\delta)$, with $\delta=0$ and $\delta=1$ leading to the ask and bid prices respectively.
With this notation, the average profits are:
\begin{equation}
\prof_T=
\int_0^T(S_t-\bar S_t)d\varphi^{\uparrow}_t+
\int_0^T(\bar S_t-(1-\varepsilon)S_t)d\varphi^{\downarrow}_t.
\end{equation}
In other words, while the investor is only sensitive to the final bid and ask prices, the market maker's profits depend separately on sales or purchases, depending on the choice of the book price.

Plugging $\bar S_t=S_t(1-\eps\delta)$ in the above expression, and passing to the limit as $T\uparrow\infty$, average profits become equal to:
\begin{equation}
\prof=\lim_{T\rightarrow\infty}\frac1T\prof_T = \eps \left(\delta\wet_{\alpha-} + \frac{1-\delta}{1-\eps}\wet_{\alpha+}\right),
\end{equation}
where $\wet_-$ and $\wet_+$ denote the expressions for expected purchases and sales, which add to absolute turnover in Theorem \ref{thm:mainresult}:
\begin{equation*}
\wet_{\alpha-}=
\frac{\sigma^2}{2}
\frac{\eta_{\alpha-}(2\bar\mu-1)}{(u(\bar\lambda)/l(\bar\lambda))^{{2\bar\mu}-1}-1}
\qquad
\wet_{\alpha+}=
\frac{\sigma^2}{2}
\frac{\eta_{\alpha+}(1-\bar\mu)}{(u(\bar\lambda)/l(\bar\lambda))^{1-2\bar\mu}-1}.
\end{equation*}

Figure \ref{fig:plotrans} shows the market maker's expected profit, as a function of the spread $\varepsilon$, for $\delta = 1, 0.5, 0$. When profits are concentrated on purchases ($\delta = 1$), the market maker optimally sets the spread in the range of 3-4\%. This value is high compared to the spreads currently observed in US and European equities, in which market makers are no longer monopolists. However, such a figure is typical of the spreads observed on small capitalization stocks \citep{amihud1986asset}, which are traded by fewer dealers.

By contrast, the model leads to unreasonably high spreads, well above 20\%, if the profits are split equally between sales and purchases, or concentrated on sales ($\delta = 0.5$ or $0$). This result, if counterintuitive initially, is implied by the asymmetry between expected sales $(\delta=0)$ and expected purchases $(\delta=1)$. Because the exponential investor wants to keep the risky position approximately fixed, and because the risky asset grows on average, the investor steadily realizes past gains over time, as the risky position reaches the selling boundary. By contrast, purchases occur after large drops in the asset price, which are much less frequent. 

This observation in fact hints at a weakness of the model, the absence of dividends, which in practice allow the investor to avoid selling shares, cashing dividends instead. Thus, using a value of $\delta$ close to $1$ is a plausible assumption, which remedies in part the absence of dividends in the model.

\subsection{Finite Horizon Bounds. Myopic Probability as Risk Neutral.}

A novel aspect of our analysis is the derivation of finite horizon bounds \citep*{guasoni.robertson.10} in the context of exponential utility. These bounds offer estimates on the performance of the candidate long-run optimal policy on any intermediate horizon $T$. They are both a mathematical device to prove the verification theorem, and a diagnostic tool to determine at which horizons the long-run optimal policy is effective enough.
For exponential utility, finite horizon bounds admit an especially appealing form -- in monetary units. As for isoelastic utilities, the respective asymptotics show that --- at the first order --- our stationary long-run policy is also optimal for any fixed finite horizon $T>0$, because the corresponding utility matches the finite horizon value function at the first nontrivial order.

\begin{theorem}[Finite horizon bounds]\label{thm:finitehorizon}
For any horizon $T>0$, the payoff $\tilde{X}_T^\phi$ of a generic admissible strategy $\phi$ in the frictionless shadow market $\tilde{S}$ satisfies
\begin{equation*}
-\frac1\alpha \log E\left[e^{-\alpha \tilde{X}^\phi_T}\right] 
\leq 
\tilde{X}^\phi_0 + \sigma^2\bar\beta T 
+\frac1\alpha \tilde E[\tilde{q}(\Upsilon_0)-\tilde{q}(\Upsilon_T)]=\tilde{X}^\phi_0 + \sigma^2\bar\beta T+O(\tfrac{\ve}{\alpha}).
\end{equation*}
For the shadow payoff corresponding to the long-run optimal strategy $\varphi$ from Theorem \ref{thm:mainresult},
\begin{equation*}
-\frac1\alpha \log E\left[e^{-\alpha \tilde{X}^\varphi_T}\right] =
\tilde{X}^\varphi_0 + \sigma^2 \bar\beta T
-\frac1\alpha \log \tilde E[e^{\tilde{q}(\Upsilon_T)-\tilde{q}(\Upsilon_0)}]=\tilde{X}^\varphi_0 + \sigma^2\bar\beta T+O(\tfrac{\ve}{\alpha}).
\end{equation*}
Here $\tilde{E}[\cdot]$ denotes the expectation with respect to the unique risk-neutral probability for $\tilde{S}$, $\Upsilon$ the centered logarithm of the risky position in Theorem \ref{thm:shadow}, and $\tilde{q}$ the deterministic function defined in Lemma \ref{lemfinite} below.
\end{theorem}

A new mathematical insight of this theorem is that, with exponential utility, the risk neutral probability replaces the myopic probability in the finite-horizon bounds. By definition, under the myopic probability the logarithmic policy coincides with the optimal policy under the real probability. Thus, strictly speaking, a risk-neutral probability is never myopic, because a logarithmic investor (in fact, any investor) takes a zero position in a risky asset with null return.

Yet, the finite-horizon bounds in the theorem above involve expectations under the risk-neutral probability, just as the similar bounds in \citet*{guasoni.robertson.10} involve expectations under the myopic probability.
The intuition is that, as relative risk aversion increases, the risky weight in the optimal isoelastic policy decreases to zero, and so does the drift under the myopic probability. Even as the weight decreases to zero, the monetary position can converge to a finite amount, as it happens for exponential utility. 

Duality theory sheds further light on the bounds. For any payoff $X$ and any stochastic discount factor (or martingale density) $M=dQ/dP$, Jensen's inequality implies that:
\begin{equation}\label{eq:duality}
-\frac1{\alpha}\log\esp{e^{-\alpha X}}=
-\frac1{\alpha}\log\esp[E_Q]{e^{-\alpha X-\log M}}\le 
\esp[E_Q]{X}+\frac1\alpha\esp[E]{M \log M}.
\end{equation}
For a fixed payoff $X$, this inequality still holds passing to the infimum over $M$, thereby indicating that equality may only hold when $M$ has minimal entropy. This abstract inequality also shows that the minimal entropy is interpreted as a monetary certainty equivalent, which represents the opportunity value of trading in the market over a given horizon. In the statement in Theorem \ref{thm:finitehorizon}, this opportunity value decomposes into the integral term, which increases linearly with the horizon, and leads to the equivalent annuity, and into the transitory term with $\tilde q$, which oscillates with the relative position of the portfolio within the no-trade region.

\subsection{Multiple Risky Assets}
Consider a market with risky assets $S^1,\ldots, S^d$ following
$$dS^i_t/S^i_t=\mu_i dt+\sigma_i dW^i_t,$$
for excess returns $\mu_i>0$, volatilities $\sigma_i>0$, and \emph{independent} standard Brownian motions $W^i$. 

With exponential utility, the optimal policy in a market with several such independent risky assets entails no-trade regions for each asset, which coincide with the no-trade regions obtained for each risky asset alone, that is, in a market with a single risky asset \citep*{liu.04}. In our setting, the following verification theorem applies, which allows to avoid the technical conditions in \citet{liu.04}. The implication is that the equivalent annuity for multiple independent risky assets is the sum of the equivalent annuities for each asset.

\begin{theorem}\label{thm:multi}
For $i=1,\ldots,d$, let $\tilde{S}^i$ be the shadow price from Lemma \ref{lem:dynamics}, with corresponding optimal strategy $(\varphi^{0,i},\varphi^i)$ from Lemma \ref{lem:ergodic} in the market with safe asset $S^0$ and risky asset $S^i$. Then, $\tilde{S}=(\tilde{S}^1,\ldots,\tilde{S}^d)$ is a shadow price in the market with safe asset $S^0$ and risky assets $S^1,\ldots,S^d$, with optimal strategy $(\sum_{i=1}^d \varphi^{0,i},\varphi^1,\ldots,\varphi^d)$ and corresponding equivalent annuity
$$\mathrm{EA}_\alpha=\sum_{i=1}^d \mathrm{EA}_\alpha^i=\sum_{i=1}^d \frac{\sigma_i^2}{2\alpha}(\bar\mu_i^2-\bar\lambda_i^2).$$
Here, $\bar\mu_i=\mu_i/\sigma_i^2$ and each gap $\bar\lambda_i$ is defined as in item $iv)$ of Theorem \ref{thm:mainresult}. Moreover, like the equivalent annuity, relative and absolute turnover also add across independent assets. 
\end{theorem}

These decompositions are unique for exponential utility, and fail for utilities in the isoelastic class. For example, \citet*{MR1372917} show that, in a market with two identical and independent assets, the no-trade region for each asset is wider than the no-trade region for a market with that asset alone. As a result, the equivalent safe rate for the two-asset market is greater than the sum of the equivalent safe rates.

\subsection{Safe Rate}
 
Throughout the paper, we assume a zero safe rate. This choice is made in part to ease notation, and the results can be adapted to the case of a constant safe rate $r$, with an important caveat. For an exponential investor with a long horizon, an arbitrarily small safe rate $r$ is preferable to \emph{any} risky investment opportunity $\bar\mu$. The reason is that the optimal policy of this particularly risk-averse investor is to keep only a bounded amount of money in the risky asset, whence her wealth on average grows linearly over time. By contrast, a positive safe rate allows wealth to grow exponentially -- without risk. Therefore, full investment in the safe asset is eventually preferred by the exponential investor in the long run. 

Nevertheless, the finite horizon bounds in Theorem \ref{thm:finitehorizon} remain valid even for a positive safe rate. Indeed, setting $\hat\alpha=e^{rT}\alpha$ in Theorem \ref{thm:finitehorizon} for \emph{discounted} payoffs, it follows that the \emph{undiscounted} payoff $\tilde X^{\phi}_T$ of any admissible strategy $\phi$ in the frictionless shadow market $\tilde S$ satisfies
\begin{equation*}
-\frac1\alpha \log E\left[e^{-\alpha \tilde{X}^\phi_T}\right] 
\leq 
e^{rT}\tilde{X}^\phi_0 + \sigma^2\bar\beta T 
+\frac1\alpha \tilde E[\tilde{q}(\Upsilon_0)-\tilde{q}(\Upsilon_T)]=e^{rT}\tilde{X}^\phi_0 + \sigma^2\bar\beta T+O(\tfrac{\ve}{\alpha T}).
\end{equation*}
Likewise, the shadow payoff of the long-run optimal strategy $\varphi$ in Theorem \ref{thm:mainresult} satisfies
\begin{equation*}
-\frac1\alpha \log E\left[e^{-\alpha \tilde{X}^\varphi_T}\right] =
e^{rT}\tilde{X}^\varphi_0 + \sigma^2 \bar\beta T
-\frac1\alpha \log \tilde E[e^{\tilde{q}(\Upsilon_T)-\tilde{q}(\Upsilon_0)}]=e^{rT}\tilde{X}^\varphi_0 + \sigma^2\bar\beta T+O(\tfrac{\ve}{\alpha T}).
\end{equation*}
Hence, our long-run optimal policy still matches the finite horizon value function up to terms of order $O(\ve/T)$. However, as the horizon becomes large, the contribution of the risky investment grows only linearly with the horizon $T$. Hence it becomes negligible, as the certainty equivalent grows exponentially.
 
\section{Heuristics}

In this section, we first use informal arguments from stochastic control to determine a candidate for the optimal policy. Then, we derive a candidate shadow price process, which is key for the subsequent verification.

\subsection{Optimal Policy}
For a trading strategy $(\varphi^0_t,\varphi_t)$, write the number of stocks $\varphi=\varphi^{\uparrow}-\varphi^\downarrow$ as the difference of the cumulative numbers of stocks purchased and sold, and denote by
$$X_t=\varphi^0_t, \quad Y_t=\varphi_t S_t,$$
the values of the safe and risky positions in terms of the ask price $S_t$. Then, the self-financing condition, and the dynamics of $S$ imply 
\begin{align*}
dX_t =& -S_t d\varphi^{\uparrow}_t+(1-\ve)S_t d\varphi^{\downarrow}_t,\\
dY_t =& \mu Y_t dt +\sigma Y_t dW_t +S_td\varphi^{\uparrow}_t-S_td\varphi_t^{\downarrow}.
\end{align*}
Consider the problem of maximizing the expected exponential utility $U(x)=-e^{-\alpha x}$ from terminal wealth at time $T$. Denote by $V(t,x,y)$ its value function, which depends on time as well as the safe and risky positions. It\^o's formula yields:
\begin{align*}
d V(t,X_t,Y_t)=& V_t dt+V_x dX_t + V_y dY_t +\tfrac 12  V_{yy} d\langle Y,Y\rangle_t\\
=& 
\left(V_t+\mu Y_t V_y+\frac{\sigma^2}2 Y_t^2 V_{yy}\right)dt\\
& +S_t(V_y-V_x)d\varphi^{\uparrow}_t+S_t((1-\ve)V_x-V_y)d\varphi^{\downarrow}_t+\sigma Y_t V_y dW_t,
\end{align*}
where the arguments of the functions are omitted for brevity. Because $V(t,X_t,Y_t)$ must be a supermartingale for any choice of the cumulative purchases and sales $\varphi^{\uparrow},\varphi^{\downarrow}$ (which are increasing processes), it follows that  $V_y-V_x \le 0$ and $(1-\ve) V_x-V_y \le 0$, that is
\begin{equation*}
1 \le \frac{V_x}{V_y}\le \frac{1}{1-\ve}.
\end{equation*}
In the interior of this region, the drift of $V(t,X_t,Y_t)$ cannot be positive, and must become zero for the optimal policy, 
\[
 V_t+\mu Y_t V_y+\frac{\sigma^2}2 Y_t^2 V_{yy} =0
 \qquad \text{if } \qquad 1< \frac{V_x}{V_y}<\frac{1}{1-\ve}.
\]
To simplify further, we use the usual scaling for exponential utility (cf., e.g., \citet*{davis.al.93}). Moreover -- in the long run -- the value function should grow exponentially with the horizon at a constant rate. This leads to the following ansatz for the value function:
 \begin{equation}\label{eq:ansatz}
 V(t,X_t,Y_t)=-e^{-\alpha X_t}  e^{\alpha \sigma^2\bar{\beta}t} v(Y_t),
 \end{equation}
which reduces the HJB equation to
\[
 \tfrac{1}{2}y^2v''(y)+\bar{\mu} y v'(y)+\alpha \bar{\beta} v(y)=0
\qquad \text{if } \qquad 1<\frac{-\alpha v(y)}{v'(y)}<\frac 1{1-\ve}.
\]
Conjecturing that the set $\{y:1<\frac{-\alpha v(y)}{v'(y)}<\frac 1{1-\ve}\}$ coincides with some interval $l<y<u$ to be determined, the following free boundary problem arises:
\begin{align}
\label{hjbred}
\tfrac{1}{2}y^2v''(y)+\bar{\mu} y v'(y)+ \alpha\bar{\beta} v(y)&=0
\qquad \text{if } l < y < u, \\
\label{boundbuy}
v'(l)+\alpha v(l)&=0,\\
\label{boundsell}
(1/(1-\ve))v'(u)+\alpha v(u)&=0.
\end{align}
These conditions are not enough to identify the solution, because they can be matched for any choice of the trading boundaries $l, u$. The optimal boundaries are the ones that also satisfy the smooth-pasting conditions (cf.\ \citet*{dumas.91}), formally obtained by differentiating \eqref{boundbuy} and \eqref{boundsell} with respect to $l$ and $u$, respectively:
\begin{align}
\label{smoothbuy}
v''(l)+\alpha v'(l)=0,\\
(1/(1-\ve))v''(u)+ \alpha v'(u)=0.\label{smoothsell}
\end{align}
In addition to the reduced value function $v$, this system requires to solve for $\bar\beta$ (and hence the equivalent annuity $\sigma^2\bar{\beta}$) as well as the trading boundaries $l$ and $u$. Substituting~\eqref{smoothbuy} and~\eqref{boundbuy}  into~\eqref{hjbred} yields 
\begin{align*}
\tfrac{1}2 (-\alpha)^2 l^2 v +\bar\mu (-\alpha)l v +\alpha \bar\beta v=0.
\end{align*}
Setting $\eta_{\alpha-}=l$, and factoring out $-\alpha v$, it follows that 
\begin{align*}
-\frac{\alpha}2 \eta_{\alpha-}^2+\bar\mu \eta_{\alpha-} -\bar\beta=0.
\end{align*}
Note that $\eta_{\alpha-}$ is the risky position when it is time to buy, and hence the risky position is valued at the ask price. 
The same argument for $u$ shows that the other solution of the quadratic equation is $\eta_{\alpha+}=u(1-\ve)$, i.e., the risky position when it is time to sell, and hence the risky position is valued at the bid price. Thus, the optimal policy is to buy when the ``ask" position falls below~$\eta_{\alpha-}$, sell when the ``bid" position rises above~$\eta_{\alpha+}$, and do nothing in between.
Since $\eta_{\alpha-}$ and $\eta_{\alpha+}$ solve the same quadratic equation, they are related to $\bar{\beta}$ via
\begin{align*}
\eta_{\alpha\pm}=\frac{\bar{\mu}}{\alpha}\pm \frac{\sqrt{\bar{\mu}^2-2 \bar{\beta}\alpha}}{\alpha}.
\end{align*}
It is convenient to set $\bar{\beta}=(\bar{\mu}^2-\bar{\lambda}^2)/2\alpha$, because $\bar{\beta}=\bar{\mu}^2/2\alpha$ without transaction costs. With this notation, the buy and sell boundaries are just
\begin{align*}\label{eq:uplow}
\eta_{\alpha\pm}=\frac{\bar{\mu}\pm\bar{\lambda}}{\alpha}.
\end{align*}
Now that $l(\bar{\lambda}),u(\bar{\lambda})$ are identified by $\eta_{\alpha\pm}$ in terms of $\bar{\lambda}$, it remains to find $\bar{\lambda}$. After deriving $l(\bar{\lambda})$ and $u(\bar\lambda)$, the boundaries in the problem \eqref{hjbred}-\eqref{boundsell} are no longer free, but fixed. With the substitution 
\begin{equation*}\label{eq:sub1}
  v(y)=e^{-\int_0^{\log(y/l(\bar\lambda))} w(z)dz}, \quad \mbox{i.e.,} \quad w(y)=-\frac{l(\bar\lambda)e^y v'(l(\bar\lambda)e^y)}{v(l(\bar\lambda)e^y)},
\end{equation*}
the boundary problem \eqref{hjbred}-\eqref{boundsell} simplifies to a Riccati ODE:
\begin{align}
w'(y)-w(y)^2+\left(2\bar\mu-1\right)w(y)-
\left(\bar\mu-\bar\lambda\right)
\left(\bar{\mu}+\bar\lambda\right)
 &= 0, \quad y\in[0,\log u(\bar\lambda)/l(\bar\lambda)], \label{riccati1}\\
w(0) &= \bar\mu-\bar\lambda,\label{riccati2}\\
w(\log (u(\bar\lambda)/l(\bar\lambda))) &= \bar\mu+\bar\lambda,\label{riccati3}
\end{align}
where
\begin{equation*}
\frac{u(\bar\lambda)}{l(\bar\lambda)} = 
\frac{1}{(1-\ve)}\frac{\eta_{\alpha+}}{\eta_{\alpha-}} =
\frac{1}{(1-\eps)}\frac{\bar\mu+\bar\lambda}
{\bar\mu-\bar\lambda}.
\end{equation*}
For each $\bar\lambda$, the initial value problem \eqref{riccati1}-\eqref{riccati2} has a solution $w(\cdot)$, which we now denote by $w(\bar\lambda,\cdot)$ with a slight abuse of notation. Thus, the correct value of $\bar\lambda$ is identified by the second boundary condition~\eqref{riccati3}:
\begin{equation}\label{lameq}
w(\bar\lambda,\log (u(\bar\lambda)/l(\bar\lambda))) =\bar\mu+\bar\lambda.
\end{equation}

\subsection{Shadow Market}
The key to making the above arguments rigorous is to find a frictionless shadow price $\tilde S$, which yields the same optimal policy as the one derived in the previous section. This step requires another heuristic argument.

As for logarithmic utility \citep*{gerhold.al.10a, gerhold.al.10b} and power utility \citepalias{gerhold.al.11}, the idea is that $\tilde S/S$, the ratio between the shadow and the ask price, should only depend on the state variable. Hence, we look for a shadow price of the form
$$\tilde{S}_t=\frac{S_t}{e^{\Upsilon_t}}g(e^{\Upsilon_t}),$$
where $e^{\Upsilon_t}=Y_t/l$ is the risky position at the ask price $S$, and centered at the buying boundary $l=\eta_{\alpha-}=(\bar\mu-\bar\lambda)/\alpha$.
The number of units $\varphi$ remains constant inside the no-trade region, so that in its interior the dynamics of $\Upsilon=\log(\varphi/l)+\log(S)$ coincides with that of $\log(S)$. Moreover, since $\Upsilon$ must remain in $[0,\log(u/l)]$ by definition, $\Upsilon$ is reflected at the boundaries. Hence, 
$$d\Upsilon_t=(\mu-\tfrac{\sigma^2}{2}) dt+\sigma dW_t +dL_t-dU_t,$$
for nondecreasing local time processes $L,U$ that only increase on $\{\Upsilon_t=0\}$ (resp. $\{\Upsilon_t=\log(u/l)\}$). The function $g:[1,u/l] \to [1,(1-\ve)u/l]$ is a $C^2$-function satisfying the smooth pasting conditions (cf. \citet*{gerhold.al.10a})
\begin{equation}\label{eq:smooth}
g(1)=1, \quad g(u/l)=(1-\ve)u/l, \quad g'(1)=1,  \quad g'(u/l)=1-\ve.
\end{equation}
The first two conditions ensure that $\tilde{S}$ equals the ask price $S$ (resp. the bid price $S(1-\ve)$) when $\Upsilon$ sits at the buying boundary $0$ (resp. at the selling boundary $\log(u/l)$), while the latter two conditions ensure that the diffusion coefficient of $\tilde S_t/S_t$ vanishes both at the bid $1-\eps$ and at the ask $1$, and hence that these bounds are not breached. The boundary conditions for $g'$, and It\^o's formula imply that $\tilde{S}$ is an It\^o process with dynamics
\begin{align*}
d\tilde{S}_t/\tilde{S}_t&=\tilde{\mu}(\Upsilon_t)dt+\tilde{\sigma}(\Upsilon_t)dW_t,
\end{align*}
where
$$\tilde{\mu}(y)=\frac{\mu g'(e^y)e^y +\frac{\sigma^2}{2}g''(e^y)e^{2y}}{g(e^y)}, \quad\text{and}\quad
\tilde{\sigma}(y)=\frac{\sigma g'(e^y)e^y}{g(e^y)}.$$
To identify the function $g$, first derive the HJB equation for a generic $g$. Then, compare this equation to the one obtained in the previous section for the market with transaction costs. Because the value function of the two problems must be the same, matching the two HJB equations identifies the function $g$.

The wealth process corresponding to a policy\footnote{Note that this is the monetary \emph{amount} rather than the fraction of wealth in the risky asset.} $\tilde{\eta}$ in terms of the shadow price $\tilde{S}$ is 
$$d\tilde{X}_t=\tilde{\eta}_t \tilde{\mu}(\Upsilon_t)dt+\tilde{\eta}_t \tilde{\sigma}(\Upsilon_t)dW_t.$$
With the standard ansatz $\tilde{V}(t,\tilde{X}_t,\Upsilon_t)$ for the value function, It\^o's formula yields
\begin{align*}
d\tilde{V}(t,\tilde{X}_t,\Upsilon_t)=&\left(\tilde{V}_t+\tilde{\mu} \tilde{\eta}_t \tilde{V}_x +\tfrac{\tilde{\sigma}^2}{2}\tilde{\eta}_t^2 \tilde{V}_{xx}+\left(\mu-\tfrac{\sigma^2}{2}\right)\tilde{V}_y+\tfrac{\sigma^2}{2}\tilde{V}_{yy}+\sigma\tilde{\sigma} \tilde{\eta}_t \tilde{V}_{xy}\right)dt\\
&+\tilde{V}_y(dL_t-dU_t)+(\tilde{\sigma}\tilde{\eta}_t \tilde{V}_x+\sigma \tilde{V}_y)dW_t,
\end{align*}
where the arguments of the functions are omitted for brevity. Since $\tilde V$ must be a supermartingale for any strategy, and a martingale for the optimal strategy, the HJB equation reads as:
$$\sup_{\tilde{\eta}}\left(\tilde{V}_t+\tilde{\mu} \tilde{\eta} \tilde{V}_x+\tfrac{\tilde{\sigma}^2}{2}\tilde{\eta}^2 \tilde{V}_{xx}+\left(\mu-\tfrac{\sigma^2}{2}\right)\tilde{V}_y+\tfrac{\sigma^2}{2} \tilde{V}_{yy}+\sigma\tilde{\sigma} \tilde{\eta} \tilde{V}_{xy}\right)=0
$$
with the Neumann boundary conditions
$$\tilde{V}_y(0)=\tilde{V}_y(\log(u/l))=0.$$
The homogeneity of the value function, (i.e., $\tilde{V}(t,x,y)=-e^{-\alpha x}\tilde{v}(t,y)$) leads to the first-order condition:
\begin{equation*}
\tilde{\eta}_t=\frac{1}{\alpha}\left(\frac{\tilde{\mu}}{\tilde{\sigma}^2}+\frac{\sigma}{\tilde{\sigma}} \frac{\tilde{v}_y}{\tilde{v}}\right).
\end{equation*}
Plugging this equality back into the HJB equation yields the nonlinear equation
$$\tilde{v}_t+\left(\mu-\frac{\sigma^2}{2}\right)\tilde{v}_y+\frac{\sigma^2}{2}\tilde{v}_{yy}-\frac{1}{2}\left(\frac{\tilde{\mu}}{\tilde{\sigma}}+
\sigma\frac{\tilde{v}_y}{\tilde{v}}\right)^2 \tilde{v}=0.$$
Now, the equivalent annuity of the optimal policy must be the same for the shadow market as for the transaction cost market in the previous section. Thus, in view of \eqref{eq:ansatz}, set
$$\tilde{v}(t,y)=e^{\alpha \sigma^2 \bar{\beta}t} e^{-\int_0^y \tilde{w}(z) dz},$$
which implies that $\tilde{v}_t=\alpha\bar\beta \sigma^2\tilde{v}$ and $\tilde{v}_y/\tilde{v}=-\tilde{w}$. Then, the HJB equation reduces to the inhomogeneous Riccati ODE
\begin{equation}\label{eq:riccati}
\tilde{w}'-\tilde{w}^2+\left(2\bar\mu-1\right)\tilde{w}-(\bar\mu^2-\bar\lambda^2)+\left(\frac{\tilde{\mu}}{\sigma\tilde{\sigma}}-\tilde{w}\right)^2=0
\end{equation}
with the boundary conditions
\begin{equation*}\label{eq:boundary}
\tilde{w}(0)=\tilde{w}(\log(u/l))=0.
\end{equation*}
For $\tilde{S}$ to be a shadow price, its value function
$$\tilde{V}_t=-e^{\alpha\sigma^2 \bar\beta t -\alpha \tilde{X}_t-\int_0^y \tilde{w}(z) dz}$$
must coincide with the value function
$$V_t=-e^{\alpha \sigma^2 \bar\beta t-\alpha X_t-\int_0^y w(z)dz}$$
for the transaction cost problem derived above. By definition, the safe position $X$ and the wealth $\tilde{X}$ in terms of $\tilde{S}=Se^{-\Upsilon}g(e^\Upsilon)$ are related via
$$\tilde{X}_t-X_t=\varphi^0_t+\varphi_t \tilde{S}_t-\varphi^0_t=g(e^{\Upsilon_t})l.$$
Now, the condition $\tilde{V}=V$ implies that
$$0=\alpha g(e^y) l+\int_0^y (\tilde{w}(z)-w(z))dz,$$
which in turn means that
\begin{equation*}\label{eq:link}
\tilde{w}(y)=w(y)-\alpha g'(e^y)e^y l.
\end{equation*}
Plugging this relation into the ODE~\eqref{eq:riccati} for $\tilde{w}$, using the ODE~\eqref{riccati1} for $w$, and simplifying gives
$$\left(-w(y)+\frac{\tilde{\mu}(y)}{\sigma\tilde{\sigma}(y)}\right)^2=0.$$
Inserting the definitions of $\tilde{\mu}(y)$ and $\tilde{\sigma}(y)$, this relation is tantamount to the following ODE for $g$:
\begin{equation}\label{eq:g1}
\frac{g''(e^y)e^{y}}{g'(e^y)}+2\bar\mu-2w(y)=0.
\end{equation}
Now, the substitution\footnote{For the second representation, we already use the boundary condition $g(1)=1$.} 
$$
k(y)=\frac{1}{g'(e^y)e^y},\quad \mbox{i.e.,} \quad g(e^y)=1+\int_0^y \frac{1}{k(z)} dz,$$
reduces this ODE to the inhomogeneous \emph{linear} equation
\begin{equation}\label{eq:k}
k'(y)=k(y)\left(2\bar\mu-1-2w(y)\right).
\end{equation}
The smooth pasting condition for $g$ implies $k(0)=1/g'(1)=1$. The solution to~\eqref{eq:k} then follows from the variation of constants formula. Plugging in the explicit formula~\eqref{eq:explfor} for $w$, and integrating, leads to (with $a=a(\bar\lambda)$ and $b=b(\bar\lambda)$ as in~\eqref{eq:ab}) the solution\footnote{Here we only show the case $\bar\mu>1/4$. The other case leads to another explicit formula, whence similar calculations follow.}
$$
k(y)=\left(1+\frac{b^2}{a^2}\right)\cos^2\left[\tan^{-1}\left(\frac{b}{a}\right)+ay\right].
$$ 
Now the chain of substitutions is reversed starting from $k$, which is known explicitly up to the constant $\bar\lambda$. First, set $\tilde{w}(y)=w(y)-\alpha l/k(y)$; then $\tilde{w}(0)=0$ by the initial conditions for $w$ and $k$. 
To establish the other boundary condition $\tilde{w}(\log(u/l))=0$, it suffices to check that $k(\log(u/l))=(\bar\mu-\bar\lambda)/(\bar\mu+\bar\lambda)$. To see this, insert the boundary condition for $w'$,
\begin{align}
\bar\mu+\bar\lambda =& w'(\log(u/l)) = \frac{a^2}{\cos^2[\tan^{-1}(\frac{b}{a})+a\log(\frac{u}{l})]},\label{eq:expl2}
\end{align}
into the explicit formula for $k(y)$.\footnote{The first equalities in~\eqref{eq:expl2} follows from the ODE for $w$, whereas the second equality is obtained from the explicit formula~\eqref{eq:explfor}.} Now, observe that the function  
\begin{align*}
g(e^y)=1+\int_0^y \frac{1}{k(z)}dz= 1+\frac{a}{a^2+b^2}\left(\tan\left[\tan^{-1}\left(\frac{b}{a}\right)+ay\right]-\frac{b}{a}\right)
\end{align*}
evidently satisfies $g(1)=1$. Moreover, $g(u/l)=(1-\ve)u/l$, which follows by inserting the terminal condition for $w$,
\begin{equation*}\label{eq:boundw}
\bar\mu+\bar\lambda=w(\log(u/l))=a\tan\left[\tan^{-1}\left(\frac{b}{a}\right)+a\log\left(\frac{u}{l}\right)\right]+\left(\bar\mu-\tfrac{1}{2}\right),
\end{equation*}
into the explicit expression for $g$. Finally, these boundary conditions for $g$ and those for $k$ imply that $g'(1)=1$ and $g'(u/l)=1-\ve$, i.e., $g$ satisfies the smooth pasting conditions~\eqref{eq:smooth} and, by construction, also the ODE~\eqref{eq:g1}. 

\section{Proofs}\label{sec:verification}

In the previous section we first used informal control arguments to find a candidate optimal policy and its corresponding value function. Then, we used this guess to derive a candidate shadow price, matching a generic shadow value function with the one of the transaction cost problem.

In this section, we prove a verification theorem for the optimal policy in the frictionless market corresponding to the candidate shadow price process, and show that the optimal shadow strategy only entails purchasing (selling) when the shadow price coincides with the ask (bid) price. Thus, the policy is also feasible and optimal in the market with transaction costs.

Key to this goal are the new finite horizon bounds for exponential utility in Theorem \ref{thm:finitehorizon}.

\subsection{Explicit formulas and their properties}

The first step to construct the shadow price is to determine, for a given small $\bar\lambda>0$, an explicit expression for the solution $w$ of the ODE~\eqref{riccati1}, complemented by the initial condition~\eqref{riccati2}.

\begin{lemma}\label{lem:riccati}
For sufficiently small $\bar\lambda>0$, the function
\begin{equation}\label{eq:explfor}
w(\bar\lambda,y)=\begin{cases} 
a(\bar\lambda)\coth[\coth^{-1}(b(\bar\lambda)/a(\bar\lambda))-a(\bar\lambda)y]+(\bar\mu-\frac{1}{2}), &\mbox{if }  \bar\mu \leq 1/4,\\
a(\bar\lambda) \tan[\tan^{-1}(b(\bar\lambda)/a(\bar\lambda))+a(\bar\lambda)y]+(\bar\mu-\frac{1}{2}), &\mbox{if } \bar\mu >1/4,
\end{cases}
\end{equation}
with
\begin{equation}\label{eq:ab}
a(\bar\lambda)=\sqrt{\Big|\bar\mu^2-\bar\lambda^2-(\tfrac{1}{2}-\bar\mu)^2\Big|} \quad \mbox{and} \quad b(\lambda)=\tfrac{1}{2}+\bar\lambda,
\end{equation}
is a local solution of 
\begin{equation}\label{eq:wode}
w'(y)-w^2(y)+\left(2\bar\mu-1\right)w(y)-(\bar\mu^2-\bar\lambda^2)=0, \quad w(0)=\bar\mu-\bar\lambda.
\end{equation}
Moreover, $y \mapsto w(\bar\lambda,y)$ is increasing in both cases.
\end{lemma}

\begin{proof}
The first part of the assertion is verified by taking derivatives. The second follows by inspection of the explicit formulas.
\end{proof}

Next, establish that the crucial constant $\bar\lambda$, which determines both the no-trade region and the equivalent annuity, is well-defined.

\begin{lemma}\label{lem:lambda}
Let $w(\bar\lambda,\cdot)$ be defined as in Lemma~\ref{lem:riccati}, and set
$$
l(\bar\lambda)=\frac{\bar\mu-\bar\lambda}{\alpha}, \quad u(\bar\lambda)=\frac{1}{(1-\ve)}\frac{\bar\mu+\bar\lambda}{\alpha}.
$$
Then, for sufficiently small $\ve>0$, there exists a unique solution $\bar\lambda$ of 
\begin{equation}\label{eq:wrbd}
  w\left(\bar\lambda,\log\left(\frac{u(\bar\lambda)}{l(\bar\lambda)}\right)\right)
  -(\bar\mu+\bar\lambda)=0.
\end{equation}
As $\ve \downarrow 0$, it has the asymptotics
\[
  \bar\lambda = \left(\frac{3}{4}\bar\mu^2\right)^{1/3} \ve^{1/3} + O(\ve).
\]
\end{lemma}
\begin{proof}
The explicit expression for~$w$ in Lemma~\ref{lem:riccati} 
implies that $w(\bar\lambda,x)$ in Lemma~\ref{lem:riccati} is analytic in both
  variables at $(0,0)$. By the initial condition in~\eqref{eq:wode}, its power series has the form
  \[
    w(\bar\lambda,y) = (\bar\mu-\bar\lambda)
      +\sum_{i=1}^\infty \sum_{j=0}^\infty W_{ij} y^i \bar\lambda^j,
  \]
  where expressions for the coefficients~$W_{ij}$ are computed by expanding the explicit expression for~$w$.
  Hence the left-hand side of the boundary condition~\eqref{eq:wrbd}
  is an analytic function of~$\ve$ and~$\bar\lambda$. Its power series
  expansion shows that the coefficients of $\ve^0\bar\lambda^{j}$ vanish for
  $j=0,1,2$, so that the condition~\eqref{eq:wrbd} reduces to
  \begin{equation}\label{eq:la eq}
    \bar\lambda^3 \sum_{i\geq0} A_i \bar\lambda^i = \ve \sum_{i,j\geq0}
      B_{ij} \ve^i \bar\lambda^j
  \end{equation}
  with (computable) coefficients~$A_i$ and~$B_{ij}$. This equation has to be solved
  for~$\bar\lambda$.
  Since
  \[
    A_0 = -\frac{4}{3\bar\mu} \qquad \text{and}
      \qquad B_{00} = \bar\mu
  \]
  are non-zero, divide the equation~\eqref{eq:la eq} by
  $\sum_{i\geq0} A_i \bar\lambda^i$, and take the third root, obtaining that, for some~$C_{ij}$,
  \[
    \bar\lambda = \ve^{1/3} \sum_{i,j\geq0} C_{ij} \ve^i \bar\lambda^j
      = \ve^{1/3} \sum_{i,j\geq0} C_{ij} (\ve^{1/3})^{3i} \bar\lambda^j \, .
  \]
  The right-hand side is an analytic function
  of~$\bar\lambda$ and~$\ve^{1/3}$, so that the implicit function
  theorem~\citep*[Theorem~I.B.4]{MR2568219}
  yields a unique solution~$\bar\lambda$ (for~$\ve$ sufficiently small),
  which is an analytic function of~$\ve^{1/3}$.
  Its power series coefficients can be computed at any order. 
\end{proof}

Henceforth, consider a small relative bid-ask spread $\ve>0$, and let~$\bar\lambda$ denote the constant in Lemma~\ref{lem:lambda}. Moreover, set $w(y):=w(\bar\lambda,y)$, $a:=a(\bar\lambda)$, $b:=b(\bar\lambda)$, and $u:=u(\bar\lambda)$, $l:=l(\bar\lambda)$. By inspection, it follows that

\begin{lemma}
In both cases of Lemma \ref{lem:riccati},
$$w'(0)=\bar\mu-\bar\lambda, \quad w'\left(\log\left(\frac{u}{l}\right)\right)=\bar\mu+\bar\lambda.$$
\end{lemma}

The next lemma states the properties of the function $k$.
\begin{lemma}
Define
$$k(y)=\begin{cases} 
\left(\frac{b^2}{a^2}-1\right)\sinh^2\left[\coth^{-1}\left(\frac{b}{a}\right)-ay\right],  &\mbox{if } \bar\mu \leq 1/4,\\
\left(\frac{b^2}{a^2}+1\right)\cos^2\left[\tan^{-1}\left(\frac{b}{a}\right)+ay\right], &\mbox{if } \bar\mu>1/4.
\end{cases}$$
Then $k$ satisfies the linear ODE
$$k'(y)=k(y)\left(2\bar\mu-1-2w(y)\right), \quad 0 \leq y \leq \log\left(\frac{u}{l}\right), $$
with boundary conditions
$$k(0)=1, \quad k\left(\log\left(\frac{u}{l}\right)\right)=\frac{\bar\mu-\bar\lambda}{\bar\mu+\bar\lambda}.$$
Moreover, $k$ is strictly decreasing and, in particular, strictly positive on $[0,\log(u/l)]$. 
\end{lemma}

\begin{proof}
That $k$ satisfies the ODE follows by insertion.
The identities $\cos^2[\tan^{-1}(x)]=1/(1+x^2)$ and
$\sinh^2[\coth^{-1}(x)]=1/(x^2-1)$ yield the boundary condition at zero, whereas the boundary condition at $\log(u/l)$ follows by inserting $w'(\log(u/l))=\bar\mu+\bar\lambda$. Finally, the ODE and a comparison argument yield that $k$ is strictly decreasing.
\end{proof}

\begin{lemma}\label{lemma:g}
For $0 \leq y \leq \log(u/l)$, define
$$g(e^y):=1+\int_0^y \frac{1}{k(z)}dz.$$
Then 
\begin{align*}
g(e^y)
       =\begin{cases} 
        1+\frac{a}{b^2-a^2}\left(\sinh\left[\sinh^{-1}\left(\frac{b}{a}\right)-ay\right]-\frac{b}{a}\right), &\mbox{if } \bar\mu \leq 1/4,\\
      1+\frac{a}{b^2+a^2}\left(\tan\left[\tan^{-1}\left(\frac{b}{a}\right)+ay\right]-\frac{b}{a}\right), &\mbox{if } \bar\mu > 1/4,
       \end{cases}
\end{align*}
and $g$ satisfies the boundary and smooth pasting conditions
$$g(1)=1, \quad g(u/l)=(1-\ve)u/l, \quad g'(1)=1, \quad g'(u/l)=1-\ve.$$
Moreover, $g'>0$ so that $g$ maps $[1,u/l]$ onto $[1,(1-\ve)u/l]$. Finally, $g$ solves the ODE
\begin{equation}\label{eq:g}
\frac{g''(e^y)e^{y}}{g'(e^y)}+2\bar\mu-2w(y)=0.
\end{equation}
\end{lemma}

\begin{proof}
The explicit representation follows by elementary integration. Evidently, $g(1)=1$. Moreover, $g(u/l)=(1-\ve)u/l$ follows by inserting $\bar\mu+\bar\lambda=w(\log(u/l))$. Next, since $g'(e^y)=1/e^{y}k(y)$, the boundary conditions for $g$ and $k$ imply the smooth pasting conditions $g'(1)=1$ and $g'(u/l)=1-\ve$. Furthermore, $k>0$ and $g'(e^y)=1/e^{y}k(y)$ show that $g'>0$. Finally, computing the derivatives verifies that $g$ indeed satisfies the ODE \eqref{eq:g}.
\end{proof}

\subsection{The shadow price and verification}

The construction of the shadow price proceeds in analogy to logarithmic utilities \citep*{gerhold.al.10a, gerhold.al.10b} and power utilities \citepalias{gerhold.al.11}. For $y \in [0,\log(u/l)]$, let $\Upsilon$ be a Brownian motion with drift, reflected at $0$ and $\log(u/l)$, that is, the continuous, adapted process with values in $[0,\log(u/l)]$ such that
\begin{equation}\label{eq:reflected}
d\Upsilon_t=(\mu-\sigma^2/2)dt+\sigma dW_t+dL_t-dU_t, \quad \Upsilon_0=y_0,
\end{equation}
for nondecreasing adapted local time processes $L$ and $U$ increasing only on the sets $\{\Upsilon_t=0\}$ and $\{\Upsilon_t=\log(u/l)\}$, respectively.

\begin{lemma}\label{lem:dynamics}
Define
\begin{equation}\label{eq:jump}
y_0=\begin{cases} 0, &\mbox{if } \xi S_0 \leq l,\\  \log(u/l), &\mbox{if } \xi S_0 \geq u,\\\log(\xi S_0/l), &\mbox{otherwise,}  \end{cases}
\end{equation}
and let $\Upsilon$ be defined as in~\eqref{eq:reflected}, started at $\Upsilon_0=y_0$. Then $\tilde{S}=Se^{-\Upsilon}g(e^\Upsilon)$, with $g$ as in Lemma~\ref{lemma:g}, is a positive It\^o process with dynamics
$$
d\tilde{S}_t/\tilde{S}_t=\tilde{\mu}(\Upsilon_t)dt+\tilde{\sigma}(\Upsilon_t)dW_t, \quad \tilde{S}_0=S_0 e^{-y_0}g(e^{y_0}),$$
for
$$
\tilde{\mu}(y)=\frac{\mu g'(e^{y})e^{y}+\frac{\sigma^2}{2}g''(e^{y})e^{2y}}{g(e^{y})}, \quad \tilde{\sigma}(y)=\frac{\sigma g'(e^{y})e^{y}}{g(e^{y})},
$$
and $\tilde{S}$ takes values in the bid-ask spread $[(1-\ve)S,S]$.
\end{lemma}

Note that the first (resp.\ second) case in~\eqref{eq:jump} occurs if the initial position $\xi S_0$ in the risky asset lies below the buying boundary $l$  or above the selling boundary $u$. Then, there is a jump from the initial position $(\varphi^0_{0-},\varphi_{0-})=(\xi^0,\xi)$, which moves the position in the risky asset to the nearest boundary of the interval $[l,u]$.  Since this initial trade involves the purchase (resp.\ sale) of shares, the initial value of $\tilde{S}$ is chosen to match the initial ask (resp.\ bid) price. 

\begin{proof}[Proof of Lemma~\ref{lem:dynamics}]
The first part of the assertion follows from the smooth pasting conditions for $g$ and It\^o's formula. As for the second part, since $g''(1) \leq 0$, a comparison argument yields that the derivative $(g'(y)y-g(y))/y^2$ of $g(y)/y$ is non-positive. Hence $g(1)/1=1$ and $g(u/l)/(u/l)=1-\ve$ yield that $\tilde{S}=Sg(e^\Upsilon)e^{-\Upsilon}$ is indeed $[(1-\ve)S,S]$-valued.
\end{proof}

The long-run optimal portfolio in the frictionless ``shadow market'' with price process $\tilde{S}$ can be determined by calculating finite horizon bounds, similarly as in \citet*{guasoni.robertson.10} for power utility. Note that for the exponential utilities considered here the myopic probability coincides with the (unique) risk-neutral probability for $\tilde{S}$.

\begin{lemma}\label{lemfinite}
$\tilde w(y)=w(y)-\alpha g'(e^y)e^y l$, with $w$ and $g$ as in Lemmas \ref{lem:riccati} and \ref{lemma:g}, solves the ODE  %\eqref{eq:riccati}-\eqref{eq:boundary}, that is:
\begin{equation}\label{hjbtil}
\tilde{w}'-\tilde{w}^2+\left(2\bar\mu-1\right) \tilde{w}-(\bar\mu^2-\bar\lambda^2)+\left(\frac{\tilde\mu}{\sigma\tilde\sigma}-\tilde{w}\right)^2=0,
\end{equation}
with boundary conditions $\tilde{w}(0)=\tilde{w}(\log(u/l))=0$. Moreover, denoting by $\tilde{q}(y)=\int_0^y \tilde{w}(z)dz$, the shadow payoff $\tilde X_T$ corresponding to the policy  $\tilde{\eta}=\frac{1}{\alpha}\left(\frac{\tilde{\mu}}{\tilde{\sigma^2}}-\frac{\sigma}{\tilde{\sigma}}\tilde{w}\right)$ (in terms of $\tilde{S}$) and the shadow discount factor $M_T=\mathcal{E}(-\int_0^\cdot \frac{\tilde{\mu}}{\tilde{\sigma}}dW_t)_T$ satisfies the following bounds:
\begin{align}
\label{primbound}
E\left[e^{-\alpha \tilde{X}_T}\right]&=e^{-\alpha \tilde{X}_0}e^{-\alpha\sigma^2 \bar\beta T} \tilde{E}\left[e^{\tilde{q}(\Upsilon_T)-\tilde{q}(\Upsilon_0)}\right]=e^{-\alpha \tilde{X}_0}e^{-\alpha\sigma^2 \bar\beta T}+O(\ve),\\
e^{-\alpha \tilde{X}_0-\tilde{E}[\log M_T]}&=e^{-\alpha \tilde{X}_0} e^{-\alpha\sigma^2 \bar\beta T}  e^{\tilde{E}[\tilde{q}(\Upsilon_T)-\tilde{q}(\Upsilon_0)]}=e^{-\alpha \tilde{X}_0}e^{-\alpha\sigma^2 \bar\beta T}+O(\ve). \label{dualbound}
\end{align}
Here, $\bar\beta=(\bar\mu^2-\bar\lambda^2)/2\alpha$ and $\tilde E[{\cdot}]$ denotes the expectation with respect to the risk-neutral probability $\tilde{Q}$ for $\tilde S$ with density process $M$.
\end{lemma}

\begin{proof}
That $\tilde{w}$ solves the ODE \eqref{hjbtil} is easily verified by taking derivatives, while the boundary conditions immediately follow from their counterparts for $w$ and $g$.

Next, note that $\tilde{\mu}, \tilde{\sigma}, \tilde{\eta}, \tilde{w}$ are functions of $\Upsilon_t$, but their argument is omitted throughout to ease notation. To prove the first bound \eqref{primbound}, notice that the shadow wealth process $\tilde X$ satisfies:
\begin{align}
e^{-\alpha\tilde{X}_T}&=e^{-\alpha \tilde{X}_0}\exp\left(-\int_0^T \alpha \tilde{\eta} \tilde{\mu}dt -\int_0^T \alpha \tilde{\eta}\tilde{\sigma}dW_t\right)\notag\\
&=e^{-\alpha \tilde{X}_0}\exp\left(\int_0^T \left(-\frac{\tilde{\mu}^2}{\tilde{\sigma}^2}+\frac{\tilde{\mu}\sigma}{\tilde{\sigma}}\tilde{w}\right)dt+\int_0^T \left(-\frac{\tilde{\mu}}{\tilde{\sigma}}+\sigma\tilde{w}\right)dW_t\right),\label{eq:wealth}
\end{align}
where the second equality follows by substituting $\tilde{\eta}=\frac{1}{\alpha}(\frac{\tilde{\mu}}{\tilde{\sigma}^2}-\frac{\sigma}{\tilde{\sigma}}\tilde{w})$. Now, It\^o's formula and the boundary conditions $\tilde{w}(0)=\tilde{w}(\log(u/l))=0$ imply
$$\tilde{q}(\Upsilon_T)-\tilde{q}(\Upsilon_0)=\int_0^T \left(\left(\mu-\frac{\sigma^2}{2}\right)\tilde{w}+\frac{\sigma^2}{2}\tilde{w}'\right)dt+\int_0^T \sigma \tilde{w} dW_t.$$
Plugging in the ODE for $\tilde{w}$, it follows that 
\begin{equation}\label{eq:q}
\tilde{q}(\Upsilon_T)-\tilde{q}(\Upsilon_0)=\int_0^T \left(-\frac{\tilde{\mu}^2}{2\tilde{\sigma}^2}+\frac{\tilde{\mu}\sigma}{\tilde{\sigma}}\tilde{w}+\alpha\sigma^2 \bar\beta\right) dt+\int_0^T \sigma\tilde{w}dW_t.
\end{equation}
Using this identity to replace $\int_0^T \sigma\tilde{w}dW_t$ in \eqref{eq:wealth} and taking expectations then yields
$$E\left[e^{-\alpha\tilde{X}_T}\right]=e^{-\alpha \tilde{X}_0}e^{-\alpha\sigma^2\bar\beta T}
%e^{-\alpha \beta T} 
E\left[M_T e^{\tilde{q}(\Upsilon_T)-\tilde{q}(\Upsilon_0)}\right].$$
The first bound now follows by noting that, since $\tilde{\mu}(\cdot)/\tilde{\sigma}(\cdot)$ is bounded on the support $[0,\log(u/l)]$ of its argument, the nonnegative local martingale $M$ is in fact a true martingale, such that $\tilde{Q}$ is well-defined.

As for the second bound, first notice that by definition of $M$ and Girsanov's theorem, 
\begin{align}\label{eq:girsanov}
 e^{-\alpha \tilde{X}_0-\tilde{E}[\log M_T]} &= \exp\left(-\alpha \tilde{X}_0+\tilde{E}\left[-\int_0^T \frac{\tilde{\mu}^2}{2\tilde{\sigma}^2}dt+\int_0^T \frac{\tilde{\mu}}{\tilde{\sigma}}d\tilde{W}_t\right]\right),
 \end{align}
 where $\tilde{W}_t=W_t+\int_0^t \frac{\tilde{\mu}}{\tilde{\sigma}}ds$ denotes a $\tilde{Q}$-Brownian motion. Again by using that the process $\tilde{\mu}/\tilde{\sigma}$ is bounded, it follows that the stochastic integral in \eqref{eq:girsanov} is a $\tilde{Q}$-martingale with vanishing expectation. Using \eqref{eq:q}, also rewritten in terms of $\tilde{W}$, to replace the Lebesgue integral in \eqref{eq:girsanov} then shows 
 $$e^{-\alpha \tilde{X}_0-\tilde{E}[\log M_T]}=e^{-\alpha \tilde{X}_0}e^{-\alpha\sigma^2\bar\beta T}\exp\left(\tilde{E}\left[\tilde{q}(\Upsilon_T)-\tilde{q}(\Upsilon_0)
+\int_0^T 
\left(\frac{\tilde{\mu}}{\tilde{\sigma}}-\sigma\tilde{w}\right)d\tilde{W}_t
\right]\right).$$
Since $\tilde{w}(\cdot)$ and $\frac{\tilde\mu(\cdot)}{\tilde\sigma(\cdot)}$ are bounded on $[0,\log(u/l)]$, the $d\tilde{W}_t$-term in this expression is a $\tilde{Q}$-martingale, which yields the second bound \eqref{dualbound}. 

The asymptotics follow by expanding the function $\tilde{q}$ as in the proof of \citetalias[Theorem 3.1]{gerhold.al.11}
\end{proof}

With the finite horizon bounds at hand, we can now establish that the policy $\tilde{\eta}$ is indeed long-run optimal in the frictionless market with price $\tilde{S}$.

\begin{lemma}\label{lem:ergodic}
The policy
\begin{equation}\label{eq:riskyfraction}
\tilde{\eta}(\Upsilon_t)=\frac{1}{\alpha}\left(\frac{\tilde{\mu}(\Upsilon_t)}{\tilde{\sigma}^2(\Upsilon_t)}-\frac{\sigma}{\tilde{\sigma}(\Upsilon_t)}\tilde{w}(\Upsilon_t)\right)=g(e^{\Upsilon_t})l
\end{equation}
is long-run optimal with equivalent annuity $\sigma^2\bar\beta$ in the frictionless market with price process $\tilde{S}$. The corresponding wealth process (in terms of $\tilde{S}$), and the numbers of safe and risky units satisfy
\begin{align*}
\tilde{X}&=(\xi^0+\xi \tilde{S}_0)+\int_0^\cdot \tilde{\eta}(\Upsilon_t) \tilde{\mu}(\Upsilon_t)dt+\int_0^\cdot \tilde{\eta}(\Upsilon_t)\tilde{\sigma}(\Upsilon_t)dW_t, \\
\varphi^0_{0-}&=\xi^0, \quad \varphi^0_t=\tilde{X}_t-\tilde{\eta}(\Upsilon_t) \quad \mbox{for }t\geq 0,\\
\varphi_{0-}&=\xi, \quad \varphi_t=\tilde{\eta}(\Upsilon_t)/\tilde{S}_t \quad \mbox{for }t\geq 0.
\end{align*}
\end{lemma}

\begin{proof}
The formulas for the trading strategy and the wealth process associated to $\tilde{\eta}$ are immediate consequences of the respective definitions. The second representation for $\tilde{\eta}$ follows by inserting the definitions of $\tilde{\mu}$, $\tilde{\sigma}$ from Lemma \ref{lem:dynamics}, the ODE \eqref{eq:g} for $g$, and $w(y)-\tilde{w}(y)=\alpha g'(e^y)e^y l$. 

Next, note that $\varphi$ is admissible for $\tilde{S}$, because \eqref{eq:riskyfraction} shows that the corresponding risky position $\tilde{\eta}$ is bounded. Now, standard duality arguments for exponential utility imply that the shadow payoff $\tilde{X}^\phi$ corresponding to \emph{any} admissible strategy $\phi$  satisfies the inequality
\begin{equation}\label{eq:db}
E\left[e^{-\alpha \tilde{X}^\phi_T}\right]\geq e^{-\alpha \tilde{X}^\phi_0-\tilde{E}[\log M_T]}.
\end{equation}
Indeed, since $\tilde{\sigma}(\Upsilon)$ is uniformly bounded and the same holds for $\phi \tilde{S}$ by admissibility of $\phi$ and $(1-\ve)S \leq \tilde{S} \leq S$, the local $\tilde{Q}$-martingale $\tilde{X}^\phi=\tilde{X}^\phi_0+\int_0^\cdot \phi_t d\tilde{S}_t$ is in fact a true $\tilde{Q}$-martingale. Now, $d\tilde{Q}|_{\mathcal{F}_T}/dP|_{\mathcal{F}_T}=M_T$ and Jensen's inequality yield
$$E\left[e^{-\alpha \tilde{X}^\phi_T}\right]=\tilde{E}\left[e^{-\alpha \tilde{X}^\phi_T-\log M_T}\right] \geq e^{-\alpha \tilde{E}[\tilde{X}^\phi_T]-\tilde{E}[\log M_T]},$$
such that \eqref{eq:db} follows from the $\tilde{Q}$-martingale property of the shadow wealth process $\tilde{X}^\phi$ . 

Inequality \eqref{eq:db} in turn yields the following upper bound, valid for any admissible strategy $\phi$ in the frictionless market with price process $\tilde{S}$:
$$
\liminf_{T \to \infty} \left(\frac{1}{-\alpha T} \log E\left[e^{-\alpha \tilde{X}^\phi_T}\right]\right) \leq \liminf_{T \to \infty} \frac{1}{T}\left(\tilde{X}^\phi_0+\frac1\alpha\tilde{E}[\log M_T]\right).
$$
The function $\tilde{q}$ in Lemma \ref{lemfinite} is bounded on the compact support of its argument $\Upsilon$. Hence, the bound \eqref{dualbound} in Lemma \ref{lemfinite} implies that the right-hand side equals $\sigma^2\bar\beta$.

Likewise, the bound \eqref{primbound} in the same lemma implies that the shadow payoff $\tilde{X}^\varphi$ (corresponding to the number of units $\varphi$, defined in terms of the policy $\tilde\eta$) satisfies, using again that $\tilde q$ is bounded,
$$
\liminf_{T \to \infty}
\frac1{\alpha T} \log E\left[e^{-\alpha \tilde X^\varphi_T}\right]=
\liminf_{T \to \infty}\left(
{\sigma^2 \bar\beta }+\frac{\tilde X^\varphi_0}T 
-\frac{1}{\alpha T}\log\tilde{E}\left[e^{\tilde{q}(\Upsilon_T)-\tilde{q}(\Upsilon_0)}\right]
\right)= \sigma^2 \bar\beta \ ,
$$
which shows that the policy $\tilde\eta$ attains this upper bound, and concludes the proof. 
\end{proof}

The next Lemma establishes that $\tilde{S}$ is a shadow price. 

\begin{lemma}\label{lem:strategy}
The number of shares $\varphi=\tilde{\eta}/\tilde{S}$ in the portfolio~$\tilde{\eta}$ in Lemma~\ref{lem:ergodic} has the dynamics  
$$d\varphi_t/\varphi_t=dL_t-dU_t.$$
Thus, $\varphi$ increases only when $\Upsilon_t=0$, that is, when $\tilde{S}$ equals the ask price, and decreases only when $\Upsilon_t=\log(u/l)$, that is, when $\tilde{S}$ equals the bid price.
\end{lemma} 

\begin{proof}
It\^o's formula applied to~\eqref{eq:riskyfraction} yields
\begin{align*}
\frac{d\tilde{\eta}(\Upsilon_t)}{\tilde{\eta}(\Upsilon_t)}=&
\frac{\mu g'(e^{\Upsilon_t})e^{\Upsilon_t}+\frac{\sigma^2}{2} g''(e^{\Upsilon_t})e^{2\Upsilon_t}}{g(e^{\Upsilon_t})}dt+\frac{\sigma g'(e^{\Upsilon_t})e^{\Upsilon_t}}{g(e^{\Upsilon_t})}dW_t+\frac{g'(e^{\Upsilon_t})e^{\Upsilon_t}}{g(e^{\Upsilon_t})}d(L_t-U_t).
\end{align*}
Integrating $\varphi=\tilde{\eta}/\tilde{S}$ by parts, inserting the dynamics of $\tilde{\eta}$ and $\tilde{S}$, and simplifying, it follows that
$$\frac{d\varphi_t}{\varphi_t}=\frac{g'(e^{\Upsilon_t})e^{\Upsilon_t}}{g(e^{\Upsilon_t})}d(L_t-U_t).$$
Since $L$ and $U$ only increase on the sets $\{\Upsilon=0\}$ and $\{\Upsilon=\log(u/l)\}$, respectively, the assertion now follows from the boundary conditions for $g$ and $g'$. 
\end{proof}

The equivalent annuity for {any} frictionless price within the bid-ask spread must be greater or equal than in the original market with bid-ask process $((1-\ve)S,S)$, because the investor trades at more favorable prices. For a \emph{shadow price}, there is an optimal strategy that only entails buying (resp.\ selling) stocks when $\tilde{S}$ coincides with the ask- resp.\ bid price. Hence, this strategy yields the same payoff when executed at  bid-ask prices, and is also optimal in the original model with transaction costs. The corresponding equivalent annuity must also be the same, since the difference due to the liquidation costs vanishes as the horizon grows in~\eqref{eq:longrun}:
 
\begin{proposition}\label{prop:shadow}
Let $\tilde{S}$ be the shadow price for $((1-\ve)S,S)$ from Lemma \ref{lem:dynamics}, and $(\varphi^0,\varphi)$ the corresponding long-run optimal strategy from Lemma \ref{lem:ergodic}. Then $(\varphi^0,\varphi)$ is long-run optimal for the bid-ask process $((1-\ve)S,S)$ as well, with the same equivalent annuity $\sigma^2\bar\beta$.
 \end{proposition}
 
  \begin{proof}  
 As $\varphi$ only increases (resp.\ decreases) when $\tilde{S}=S$ (resp.\ $\tilde{S}=(1-\ve)S$), the strategy $(\varphi^0,\varphi)$ is self-financing for the bid-ask process $((1-\ve)S,S)$ as well. Moreover, \eqref{eq:riskyfraction} and $(1-\ve)S \leq \tilde{S} \leq S$ show that it is also admissible for $((1-\ve)S,S)$ in the sense of Definition \ref{defi:admissible}.

 Now, since $S \geq \tilde{S} \geq S(1-\ve)$ and the number $\varphi_t$ of shares is always positive and bounded from above by $u/S>0$, 
\begin{equation*}
\varphi^0_t+\varphi_t \tilde{S}_t \geq \varphi^0_t +\varphi_t^+(1-\ve)S_t-\varphi^-_t S_t  \geq \varphi^0_t+\varphi_t \tilde{S}_t-\ve u.
\end{equation*}
These upper and lower bounds yield:
\begin{equation}\label{eq:optimalrate}
\liminf_{T \to \infty} \frac{1}{-\alpha T}\log E\left[e^{-\alpha (\varphi^0_T+\varphi_T^+ (1-\ve)S_T -\varphi_T^- S_T)}\right]
 = \liminf_{T \to \infty} \frac{1}{-\alpha T}\log E\left[e^{-\alpha(\varphi^0_T+\varphi_T\tilde{S}_T)}\right],
\end{equation}
that is, $(\varphi^0,\varphi)$ has the same growth rate, either with $\tilde{S}$ or with $[(1-\ve)S,S]$. 

Now let $(\psi^0,\psi)$ be any admissible strategy for the bid-ask spread $[(1-\ve)S,S]$, and define the corresponding cash position in the shadow market as  $\tilde{\psi}^0=\psi^0_{0-}-\int_0^{\cdot} \tilde{S}_t d\psi_t.$ Then $(\tilde{\psi}^0,\psi)$ is a self-financing trading strategy for the shadow price $\tilde{S}$, and $\tilde{\psi}^0 \geq \psi^0$ because $(1-\ve)S \leq \tilde{S} \leq S$ implies that $\int_0^{\cdot} \tilde{S}_t d\psi_t \le \int_0^{\cdot} {S}_t d\psi^\uparrow_t - (1-\ve)\int_0^{\cdot} {S}_t d\psi^\downarrow_t$. Together with $\tilde{S} \in [(1-\ve)S,S]$, the long-run optimality of $(\varphi^0,\varphi)$ for $\tilde{S}$, and~\eqref{eq:optimalrate}, it follows that:
\begin{align*}
\liminf_{T \to \infty} \left( \frac{1}{-\alpha T}\log E\left[e^{-\alpha(\psi^0_T+\psi_T^+ (1-\ve)S_T -\psi_T^- S_T)}\right]\right)
&\leq \liminf_{T \to \infty} \left( \frac{1}{-\alpha T}\log E\left[e^{-\alpha(\tilde{\psi}^0_T+\psi_T \tilde{S}_T)}\right]\right)\\
\leq \liminf_{T \to \infty} \left(\frac{1}{-\alpha T}\log E\left[e^{-\alpha(\varphi^0_T+\varphi_T \tilde{S}_T)}\right]\right)
&= \liminf_{T \to \infty} \left(\frac{1}{-\alpha T}\log E\left[e^{-\alpha(\varphi^0_T+\varphi_T^+ (1-\ve)S_T -\varphi_T^- S_T)}\right]\right).
\end{align*}
Hence $(\varphi^0,\varphi)$ is also long-run optimal for the bid-ask process $((1-\ve)S,S)$.
\end{proof}

Putting everything together, we can now complete the proofs of our main results:

\begin{proof}[Proofs of Theorem \ref{thm:mainresult} $i)-iv), vi)$, Theorem \ref{thm:shadow}, Theorem \ref{thm:finitehorizon}]
By Lemma~\ref{lem:ergodic}, the strategy $(\varphi^0,\varphi)$ is optimal in the frictionless market with price process $\tilde{S}$. Since the latter is a shadow price process by Lemma~\ref{lem:strategy}, Proposition~\ref{prop:shadow} yields that the same strategy is also optimal with the same equivalent annuity $\sigma^2\bar\beta=\frac{\sigma^2}{2\alpha}(\bar\mu^2-\bar\lambda^2)$ in the original market with transaction costs. This proves Theorem \ref{thm:shadow} and also Item $i)$ of Theorem \ref{thm:mainresult}, since the definition of $\bar\lambda$ in Lemma \ref{lem:riccati} matches $iv)$ by Lemma \ref{lem:lambda}. Item $ii)$ of Theorem \ref{thm:mainresult} follows immediately by comparing the growth rate to its frictionless value. Next, since $\tilde{S}$ is a shadow price, the buy resp.\ sell boundaries for $(\varphi^0,\varphi)$ are quoted in terms of the ask resp.\ bid price. Item $iii)$ then follows from the representation in Lemma \ref{lem:ergodic}, combined with the boundary conditions for $g$ and the definitions of $u, l$ in Lemma~\ref{lem:lambda}. The corresponding asymptotic expansions in $vi)$ are an immediate consequence of the fractional power series for~$\bar\lambda$ (cf.\ Lemma~\ref{lem:lambda}) and Taylor expansion. Finally, Theorem \ref{thm:finitehorizon} has been established in Lemma \ref{lemfinite} and the proof of Lemma \ref{lem:ergodic}.
\end{proof}

Next, we prove Theorem \ref{thm:multi}, which generalizes the finite-horizon bounds to a market with several uncorrelated assets.

\begin{proof}[Proof of Theorem \ref{thm:multi}]
Let $M^i=\mathcal{E}(-\int_0^\cdot \frac{\tilde{\mu}_i(\Upsilon^i_t)}{\tilde{\sigma}_i(\Upsilon^i_t)}dW^i_t)$ be the stochastic discount factor in the market $(S^0,\tilde{S}^i)$, where the coefficients $\tilde\mu_i,\tilde\sigma_i$ and the reflected Brownian motions $\Upsilon^i$ are defined as in Lemma \ref{lem:dynamics}.
Then, since the risky assets $S^i$ are independent, the same holds for the processes $\Upsilon^i$, $M^i$, $\tilde{S}^i$. For the shadow wealth process $\tilde{X}^\varphi = \tilde{X}^\varphi_0+\sum_{i=1}^d \int_0^\cdot \varphi^i_t d\tilde{S}^i_t$, the first univariate finite horizon bound in Lemma \ref{lemfinite} therefore yields
\begin{equation*}
-\frac1\alpha\log
E\left[e^{-\alpha \tilde{X}^\varphi_T}\right]=
\tilde{X}^\varphi_0 +\sum_{i=1}^d \sigma_i^2 \bar\beta_i T-\frac{1}{\alpha}\sum_{i=1}^d \log \tilde{E}_i \left[e^{\tilde{q}_i(\Upsilon^i_T)-\tilde{q}_i(\Upsilon^i_0)}\right],
\end{equation*}
where $\tilde{E}_i[\cdot]$ denotes expectation with respect to the measure with density process $M_i$, and the constants $\bar\beta_i$ and the functions $\tilde{q}_i$ are defined as in Lemma \ref{lemfinite} for $i=1,\ldots,d$. Since the mappings $\tilde{q}_i$ are bounded on the compact supports of the $\Upsilon^i$, it follows that
\begin{equation}\label{eq:upper}
\liminf_{T \to \infty}-\frac{1}{\alpha T}\log E\left[e^{-\alpha X^{\varphi}_T}\right]=\sum_{i=1}^d \sigma_i^2\bar\beta_i.
\end{equation}
 Next, notice that since each $\tilde{S}^i$ only depends on one of the independent Brownian motions, Girsanov's theorem implies that $M=\mathcal{E}(-\sum_{i=1}^d \int_0^\cdot \frac{\tilde{\mu}_i(\Upsilon^i_t)}{\tilde{\sigma}_i(\Upsilon^i_t)}dW^i_t)$ is a stochastic discount factor for the market $(S^0,\tilde S^1,\ldots,\tilde S^d)$. Hence it follows verbatim as in the proof of Lemma \ref{lem:ergodic} that the shadow wealth process $\tilde{X}^\phi$ associated to any admissible strategy $\phi$ satisfies
 $$E\left[e^{-\alpha \tilde{X}^\phi_T}\right]\geq e^{-\alpha \tilde{X}^\phi_0-\tilde{E}[\log M_T]},$$
 where $\tilde{E}[\cdot]$ denotes the expectation with respect to the measure with density process $M$. Since each of the $M^i$ only depends on one of the independent Brownian motions, Yor's formula implies that $M_T=\prod_{i=1}^d M^i_T$ and hence, by independence of the $M^i$,
 $$E\left[e^{-\alpha \tilde{X}^\phi_T}\right]\geq e^{-\alpha \tilde{X}^\phi_0} \prod_{i=1}^d e^{-\tilde{E}_i[\log M^i_T]},$$
 where $\tilde{E}_i[\cdot]$ denotes the expectation with respect to the measure with density process $M^i$. Combined with the second univariate finite horizon bound from Lemma \ref{lemfinite}, it follows that
  $$E\left[e^{-\alpha \tilde X^{\phi}_T}\right] \geq  e^{-\alpha \tilde{X}^\phi_0} \prod_{i=1}^d  e^{-\alpha\sigma_i^2\bar\beta_i T+\tilde{E}_i[\tilde{q}_i(\Upsilon^i_T)-\tilde{q}_i(\Upsilon^i_0)]}.$$
In view of the boundedness of the $\tilde{q}_i$, this inequality yields
 $$  \liminf_{T \to \infty}-\frac{1}{\alpha T}\log E\left[e^{-\alpha \tilde X^{\phi}_T}\right] \leq \sum_{i=1}^d \sigma_i^2\bar\beta_i.$$
 Together with \eqref{eq:upper}, it follows that the strategy $\varphi$ is optimal in the frictionless market with risky asset $\tilde{S}$. Since, by definition of $\tilde{S}^i$ and Lemma \ref{lem:strategy}, the strategy $\varphi$ only purchases resp.\ sells shares of the risky asset $i$ when $\tilde{S}^i=S^i$ resp.\ $\tilde{S}^i=(1-\varepsilon)S^i$, the process $\tilde{S}$ is a shadow price. It then follows as in the proof of Proposition \ref{prop:shadow} that the same strategy is also optimal with the same equivalent annuity in the original market with transaction costs, completing the proof.
 \end{proof}

\subsection{Trading volume}

As above, let $\varphi_t=\varphi_t^{\uparrow}-\varphi_t^{\downarrow}$ denote the number of risky units at time $t$, written as the difference of the cumulated numbers of shares bought resp.\ sold until $t$. \emph{Relative turnover}, defined as the measure $\tilde S_t d\|\varphi\|_t/\tilde S_t\varphi_t=d\|\varphi\|_t/\varphi_t=d\varphi_t^{\uparrow}/|\varphi_t|+d\varphi^{\downarrow}_t/|\varphi_t|$, is a scale-invariant indicator of trading volume, compare \citet*{lo2000trading}.
The \emph{long-term average relative turnover} is defined as
$$
\lim_{T\rightarrow\infty}\frac1T\int_0^T \frac{d\|\varphi\|_t}{|\varphi_t|}.
$$
Similarly, \emph{absolute turnover} $(1-\ve)S_td\varphi^{\downarrow}_t+S_t d\varphi^{\uparrow}_t$ is defined as the amount of wealth traded, evaluated in terms of the bid price $(1-\ve)S_t$ when selling resp.\ in terms of the ask price $S_t$ when buying. As above, the \emph{long-term average absolute turnover} is then defined as
$$
\lim_{T\rightarrow\infty}\frac1T\left(\int_0^T (1-\ve)S_td\varphi^{\downarrow}_t+\int_0^T S_t d\varphi^{\uparrow}_t\right).
$$

These quantities can be expressed in terms of the long-run averages of the local times $L$ and $U$:

\begin{proposition}
The long-term average relative turnover is 
\begin{align*}
\lim_{T\rightarrow\infty}\frac1T\int_0^T \frac{d\|\varphi\|_t}{|\varphi_t|}=\lim_{T\to \infty} \frac{U_T}{T}+\lim_{T\to \infty} \frac{L_T}{T}.
\end{align*}
The long-term average absolute turnover is
\begin{align*}
&\lim_{T\to \infty} \frac{1}{T}\left(\int_0^T (1-\ve)S_td\varphi^{\downarrow}_t+\int_0^T S_t d\varphi^{\uparrow}_t\right) =\frac{\bar\mu+\bar\lambda}{\alpha}\lim_{T\to \infty} \frac{U_T}{T}+\frac{\bar\mu-\bar\lambda}{\alpha}\lim_{T\to \infty} \frac{L_T}{T}.
\end{align*}
\end{proposition}

\begin{proof}
The formula for the relative turnover follows form the representation for $d\varphi/\varphi$ in Lemma~\ref{lem:strategy}. The formulas for the absolute turnover follow analogously by noting that $S_t\varphi_t=(\bar\mu-\bar\lambda)/\alpha$ on the set $\{\Upsilon_t=0\}$ where $L$ increases, and likewise $(1-\ve)S_t\varphi_t=(\bar\mu+\bar\lambda)/\alpha$ on the set $\{\Upsilon_t=\log(u/l)\}$ where $U$ increases. 
\end{proof}

Using the long-term limits of the local times $L$ and $U$ determined in \citetalias[Lemma D.2]{gerhold.al.11}, it follows that the long-run averages of the local times admit explicit formulas in terms of the gap $\bar\lambda$. These in turn yield the asymptotic expansions for $\ve \downarrow 0$ stated in Theorem \ref{thm:mainresult} via Taylor expansion.

\subsection{Connection to constant relative risk aversion}

Finally, we prove Theorem \ref{thm:raa}, which states that all relevant quantities, i.e., liquidity premium, optimal policy, equivalent annuity, and trading volume, for an investor with constant \emph{absolute} risk aversion arise in the limit for increasing constant \emph{relative} risk aversion.

To this end it suffices to show that the gap of \citepalias[Lemma B.2]{gerhold.al.11} for relative risk aversion $\gamma$ converges to its counterpart in our Lemma \ref{lem:lambda} for constant absolute risk aversion $\alpha$, as $\gamma \uparrow \infty$. Note that this convergence holds for any level of absolute risk aversion, since our gap is independent of the latter.

\begin{theorem}\label{thm:convergence}
As the relative risk aversion $\gamma$ in \citepalias[Lemma B.2]{gerhold.al.11} tends to infinity, their gap $\bar\lambda_\gamma$ converges to our counterpart $\bar\lambda$ in Lemma \ref{lem:lambda}, which is the gap for all levels $\alpha$ of absolute risk aversion.
\end{theorem}

\begin{proof}
For small $\ve$, the gap\footnote{In \citetalias[Appendix B]{gerhold.al.11}, this is the gap in business time, obtained by their formulas with $\mu$ and $\sigma$ replaced by $\bar\mu$ and $1$, respectively.}  $\bar\lambda_\gamma$ of \citetalias{gerhold.al.11} is given by the unique root of the function 
$$f_\gamma(\bar\lambda)=\gamma w_\gamma(\bar\lambda,\log[u_\gamma(\bar\lambda)/l_\gamma(\bar\lambda)])-(\bar\mu+\bar\lambda),$$
where the function $w_\gamma$ is given explicitly in \citetalias[Lemma B.1]{gerhold.al.11}. Now note that as the relative risk aversion $\gamma$ becomes large, Case 2 of \citetalias[Lemma B.1]{gerhold.al.11} applies if $\bar\mu>1/4$ and Case 3 applies if $\bar\mu \leq 1/4$. By inspection of the explicit formulas in \citetalias[Lemma B.1]{gerhold.al.11} resp.\ our Lemma \ref{lem:riccati}, it follows that, as $\gamma \uparrow \infty$, the function $\gamma w_\gamma(\cdot)$ converges uniformly on compacts to $w$ from our Lemma \ref{lem:riccati}. Since the same holds for the functions $u_\gamma(\cdot), l_\gamma(\cdot)$ from \citetalias[Lemma B.2]{gerhold.al.11} and $u(\cdot), l(\cdot)$ from our Lemma \ref{lem:lambda}, this in turn yields that $f_\gamma(\cdot)$ converges uniformly on compacts to 
$$f(\bar\lambda)=w(\bar\lambda,\log[u(\bar\lambda)/l(\bar\lambda)])-(\bar\mu+\bar\lambda).$$
Since our gap $\bar\lambda$ is the unique root of this function, it suffices to show that the zeros of $f_\gamma$ also converge as $\gamma \uparrow \infty$. But this follows, because a calculation shows that, for small $\ve$, the derivative $\frac{\partial}{\partial \bar\lambda}f$ is bounded away from zero in a neighborhood of the root of $f$, completing the proof.
\end{proof}

The convergence of all other -- suitably rescaled -- quantities follows immediately from the explicit formulas in \citetalias{gerhold.al.11} and in this paper.

\bibliographystyle{agsm}
\bibliography{tractrans}

\end{document}